\documentclass[amsmath,amssymb,onecolumn,superscriptaddress]{revtex4}

\usepackage{graphicx}% 
\usepackage{dcolumn}% 
\def\be{\begin{equation}}
\def\ee{\end{equation}}
\def\beq{\begin{eqnarray}}
\def\eeq{\end{eqnarray}}

\usepackage{bm}
\usepackage{graphicx}
\usepackage{dcolumn}
\usepackage{amsmath}
\usepackage[latin1]{inputenc}
\usepackage{graphicx, psfrag}
\usepackage{amssymb}
\usepackage[colorlinks=true, citecolor=blue, urlcolor = blue, linkcolor= red, bookmarks=true]{hyperref}
\usepackage{float}
\usepackage{amsmath}
\usepackage{amsfonts}
\usepackage{dcolumn}
\usepackage{hyperref}
\usepackage{subfigure}
\usepackage{pgfplots}
\usepackage{epstopdf}
\usepackage{booktabs}
\usepackage{orcidlink}
\usepackage{xcolor}

\begin{document}
\title{Well behaved class of Heintzmann's solution within $f(R,\,T)$ framework}

\author{Pramit Rej \orcidlink{0000-0001-5359-0655} \footnote{Corresponding author}}
\email[Email:]{pramitrej@gmail.com, pramitr@sccollegednk.ac.in, pramit.rej@associates.iucaa.in}
 \affiliation{Department of Mathematics, Sarat Centenary College, Dhaniakhali, Hooghly, West Bengal 712 302, India}

\author{Akashdip Karmakar \orcidlink{0009-0007-3848-1443}}
\email[Email:]{akashdip999@gmail.com}
 \affiliation{Department of Mathematics,  Indian Institute of Engineering Science and Technology, Shibpur, Howrah, West Bengal 711 103, India.}

\begin{abstract}\noindent

The primary objective of this paper is to develop a well-behaved class of Heintzmann IIa [{\em H. Heintzmann, Z. Physik 228, 489-493 (1969)}] solution in the context of $f(R,\, T)$ gravity. In the $f(R, T)$ framework, the gravitational action includes both the Ricci scalar ($R$) and the trace of the energy-momentum tensor ($T$). We chose a particular $f(R,\,T)$ model s.t. $f(R,\,T) = R+2 \chi T$, where $\chi$ is known as the coupling parameter.
This solution describes a novel isotropic compact fluid sphere with positively finite central pressure and density in this extended theory of gravity. The results obtained analytically are better described by graphical representations of the physical parameters for various values of the coupling parameter $\chi$. The solution for a specific compact object, Vela X-1, with radius $\mathfrak{R} = 9.56_{-0.08}^{+0.08}$ km and mass $\mathcal{M} = 1.77 \pm 0.08~\mathcal{M}_{\odot}$ [{\em M. L. Rawls et al. ApJ, 730, 25 (2011)}], is shown here. We analyze the fundamental physical attributes of the star, which reveals the influence of the coupling parameter $\chi$ on the values of substance parameters. This helps us to make a fruitful comparison of this modified $f(R,\, T)$ gravity with the standard GR and notice that it holds good for stable compact objects. In this framework, the star under our consideration exhibits a stable structure consistent with the Heintzmann IIa {\em ansatz}. From all of our obtained graphical and numerical results, we can ultimately conclude that our reported model is physically admissible and satisfies all the physical criteria for an acceptable model.

\end{abstract}

\maketitle
\textbf{Keywords:} Isotropic; Ricci scalar; Heintzmann IIa solution; $f(R,\,T)$ gravity; TOV equation.

\section{Introduction}

The creation of truly compact configurations that satisfy the Einstein field equations (EFE) under a broad range of physical conditions remains a major challenge. The study of compact star structures in relativistic astrophysics depends significantly on precise solutions to the EFEs, and that is why the goal of astrophysicists has been to model superdense objects. In 1916, Schwarzschild developed the first solution to Einstein's field equations \cite{schwarzschild1916gravitationsfeld1}. This suggested solution shows the vicinity of a static, compact, spherically symmetric stellar object with vanishing pressure and density.
Gravitational collapse, neutron stars, black holes, and quark stars are examples of mechanisms that belong to the class of systems with spherical symmetry, which makes spherically symmetric problems crucial to general relativity. In this context, the analysis of static configurations with spherical symmetry composed of isotropic pressure distributions and ideal fluid distributions (i.e. $p_r = p_t$) is the most straightforward scenario. Subsequently, to make the resulting set of field equations easier to solve, Tolman suggested adding an extra equation that is required to give a determinate problem in the form of an ad hoc relation between the metric tensor components, rather than using the equation of state of matter to close the system of the field equations \cite{tolman1939static}. This technique was used to obtain eight solutions of the field equations, and it is currently regarded as one of the key techniques for locating precise interior solutions of the gravitational field equations for fluid spheres. In this connection, the renowned constraint relation on the mass-to-radius ratio for stable general relativistic spheres, $\frac{2G\mathcal{M}}{c^2\mathfrak{R}} \leq \frac{8}{9}$ was obtained by Buchdahl, who made a significant contribution to the research of fluid spheres (where $G$ is the gravitational constant, $\mathcal{M}$ is the object mass, $\mathfrak{R}$ is the radius of the star and $c$ is the speed of light). 
Based on a specific selection of the mean density within the star, an exact non-singular solution was also acquired. \\
In this connection, it should be mentioned that scientists are stumped by the idea that the universe is growing faster than expected because of some sort of secret energy that has existed since the Big Bang. There may be a link between large-scale alteration of gravity and this late-time accelerated expansion of the cosmos \cite{riess1998observational,koyama2016cosmological}. Both the acceleration of expansion and the modification of gravity compared to General Relativity (GR), are caused actually by the presence of dark energy (DE), dark matter(DM), or a new form of matter, and interestingly these two scenarios are not in any way related, leading to numerous stellar models \cite{copeland2006dynamics, joyce2015beyond}. 
\textcolor{blue}{Dark energy is a fictitious form of energy, which in its most basic form is a cosmological constant; however, other possibilities include a breakdown of GR on large scales or the effect of explaining the observation using a metric not accurate for our inhomogeneous Universe. Right now, the $\Lambda$CDM paradigm-which holds that there are three massless neutrinos, the universe is spatially flat, and a cosmological constant dominates its energy budget-is the most practical framework for explaining all of the observations regarding the accelerated expansion of the Universe that are currently accessible. The $\Lambda$CDM model describes the evolution of the Universe with minimal cosmological parameters. Current data constrains these values at the $\%$ level. A variety of probes, each with pros and cons, can be used to place limits on cosmological parameters, particularly dark energy parameters. The most well-known of these are the Cosmic Microwave Background (CMB), Baryonic Acoustic Oscillation (BAO), Supernovae type Ia (SNe), as well as probes of the growth of structure through weak lensing investigations and cluster of galaxies abundance. The WMAP team's article series \cite{WMAP:2003pyh, WMAP:2010qai} are the most useful resources that provides an overview of integrating several probes to derive cosmological parameters.}
Several researchers have suggested various modified theories of gravity, such as $f(R)$-gravity, $f(Q, T)$-gravity, $f(\mathcal{G}, T)$-gravity, $f(R, \mathcal{G})$-gravity, unimodular gravity, teleparallel gravity, etc. \cite{nojiri2016unimodular, garcia2019cosmic, nojiri2007introduction, nojiri2011unified, xu2019regular, arora2020f, bahamonde2023teleparallel, atazadeh2014energy, de2011stability}. where $R$ indicates the Ricci scalar, $T$ denotes the trace of the stress-energy tensor and $\mathcal{G}$ indicates the Gauss-Bonnet invariant.
T. Harko and his associates proposed $f(R, T)$ gravity theory in 2011 \cite{harko2011f}. A generalization of the gravitational Lagrangian of the $f (R)$ gravity serves as its foundation. An arbitrary function of the Ricci scalar $R$ and the trace of the energy-momentum tensor $T$ determine the gravitational Lagrangian of the standard Hilbert-Einstein action in the $f(R, T)$ theory of gravity. \\
Such a reliance on $T$ could result from taking into account quantum effects (conformal anomaly) or from an exotic imperfect fluid. The $f(R, T)$ theory of gravity, which is predicated on the non-minimal curvature matter coupling, can be regarded as a helpful formulation among the other modified gravity theories.
Without adding new spatial dimensions or introducing an exotic dark energy component, this $f(R, T)$ gravity offers an alternate explanation for the current cosmic acceleration. Due to the matter and geometry coupled together, this gravity model relies on a source term that is nothing more than the variation of the matter stress-energy tensor. \\
Consequently, the motion of the test particle deviates from the geodesic path because there is an additional force perpendicular to the four-velocity directions. In this updated $f(R, T)$ hypothesis, the cosmic acceleration arises from both the matter content and the geometrical aspect \cite{chakraborty2013alternative}. The shift from a matter-dominated era to an accelerated phase is explored by Houndjo \cite{doi:10.1142/S0218271812500034}, who picked $f(R, T)$ as $f_1(R) + f_2(T)$, where $f_1(R)$ is a function of the Ricci scalar and the $f_2(T )$ can be considered as a matter correction term to $f(R)$ gravity. The trace of energy-momentum tensor disappears for ultrarelativistic fluids; consequently, these constituents of matter do not affect the function of $f(R, T)$. To address this shortcoming, a generalization of this theory has also been developed, incorporating a new invariant, or $R_{\alpha\beta}T^{\alpha\beta}$ \cite{sharif2013energy, odintsov2013f, PhysRevD.90.044031}. Functions of $f(R, T)$ that are distinguishable either minimally or non-minimally, into arbitrary functions of the trace of the energy-momentum tensor, $h(T)$, and an arbitrary function of the Ricci scalar, $g(R)$, like $g(R) + h(T)$, $g(R)h(T)$, $g(R)\{1 + h(T)\}$, etc. \textcolor{blue}{Later found that, the cosmological solution to the second type of pure non-minimal assumption $g(R)h(T)$ is not physically justified \cite{Shabani:2013djy, Shabani:2014xvi}.}
\\
\textcolor{blue}{Several fields have already witnessed impressive outcomes from the $f(R, T)$ gravity. Such as application of $f(R, T)$ gravity for massive pulsars by Santos Jr. et al \cite{dos2019conservative}, Moraes et al \cite{moraes2016stellar}; cosmological consequences without dark energy by Sun and Huang \cite{sun2016cosmology}; dark matter theories by Zaregonbadi et al \cite{zaregonbadi2016dark}; investigations on wormholes by Moraes and his co-workers \cite{moraes2019wormholes, moraes2019charged, moraes2018nonexotic}, Elizalde and Khurshudyan \cite{elizalde2018wormhole}, Sahoo et al \cite{sahoo2020wormhole}; study on gravitational waves by Sharif and Siddiqa \cite{sharif2019propagation}, Alves et al \cite{alves2016gravitational}; bouncing cosmological scenario by Sahoo et al \cite{sahoo2020bouncing}, Bhattacharjee and Sahoo \cite{bhattacharjee2020redshift}; and so on.}
In their discussion of the non-static spherically symmetric line element, Sharif et al. addressed the stability of a collapsing spherical body of an isotropic fluid distribution \cite{sharif2014dynamical}. However, to solve the instability problem, Noureen et al. created a perturbation scheme for determining the collapse equation and a constraint on the adiabatic index for the Newtonian and post-Newtonian periods \cite{noureen2015dynamical} and additionally, the range of instability for an anisotropic background confined by zero expansion under the $f(R, T)$ theory has been developed by them \cite{noureen2015dynamical1}. Using the analytic solution of the Krori and Barua metric to the spherically symmetric anisotropic star, Zubair et al. examined the prospect of the development of compact stars in the $f(R, T)$ theory of gravity \cite{zubair2016possible}.
The spherical equilibrium configuration of strange and polytropic stars under the $f(R, T)$ theory of gravity has been studied by Moraes et al \cite{moraes2016stellar}. Ahmed et al. examined how gravitational lensing is affected by $f(R, T)$ gravity and contrasted their findings with the standard outcome of GR \cite{alhamzawi2016gravitational}. \par

\textcolor{blue}{In this work, we have considered the well-known Heintzmann IIa {\em ansatz} for the particular $f(R, T)= R+2 \chi T$ framework, where $\chi$ is some arbitrary constant to explain a static, spherically symmetric arrangement connected to an isotropic fluid substance  that represents compact spherical objects. 
In 1969, H. Heintzmann \cite{heintz, Delgaty:1998uy} developed a method to derive new exact EFE solutions for an ideal isotropic fluid from the solutions obtained previously. Later, researchers discovered that it could be useful in a variety of astrophysical experiments \cite{andrade2022anisotropic, pradhan2015anisotropic}. Several approaches have already been used to expand this solution and produce physically intriguing results. Heintzmann's solution is so simple, making it easy to study general relativistic stars. To create new static and spherical solutions for isotropic fluid distributions, it acts as an initial solution. This is essential to comprehend the interior of relativistic compact objects, which might display variations from isotropy as a result of phase transitions, high densities, or strong magnetic fields. Moreover, The framework of gravitational decoupling, a technique to expand the solution space of Einstein's field equations by taking into account more sources of gravity, makes use of the Heintzmann IIa solution and so it facilitates the investigation of more complicated compact object models. The framework of the Heintzmann IIa solution includes both the trace of the energy-momentum tensor and the Ricci scalar and so, compared to models that might only take one of these aspects into account, this dual inclusion enables a more thorough understanding of gravitational interactions. Assuming isotropic pressure inside the fluid sphere is the foundation of the Heintzmann IIa solution. Anisotropic pressure can, however, actually exist inside compact objects for a variety of reasons, including rotation, phase transitions, and strong magnetic fields. So, this could reduce how accurately the solution reflects actual objects. Overall, the Heintzmann IIa solution provides an innovative viewpoint on the stability and internal structure of compact objects in the universe and is a strong model inside the $f(R, T)$ framework.}
Pant et al. \cite{Pant:2010byj} introduced a well-behaved class of charged analogs of Heintzmann's relativistic solutions. Thirukkanesh et al. \cite{Thirukkanesh:2018hfy} presented a generalized algorithm for EFE solutions that can be reduced to the Heintzmann IIa solution. Anisotropy and charge were added to the Heintzmann solution by Singh and Pant \cite{Singh:2015kyr} to simulate dense compact objects. By using the gravitational decoupling technique, Estrada and Tello-Ortiz \cite{Estrada:2018zbh} were able to obtain the new anisotropic solution by using this solution as an interior solution. Morales and Tello-Ortiz \cite{Morales:2018nmq} developed charged anisotropic solutions to the EFEs by extending the Heintzmann solution by utilizing gravitational decoupling. Zubair et al. \cite{Zubair:2021zqs} extended the well-known charged isotropic Heintzmann solution to its anisotropic domain by using the gravitational decoupling approach. This Heintzmann solution has been taken into consideration in several other research projects \cite{Sharif:2020lbt, Sharif:2019zyh, Andrade:2022idg} (and further references therein).\par
\textcolor{blue}{For this investigation, we have chosen a particular compact star candidate Vela X-1, which is a high-mass X-ray binary (HMXB) comprising a massive companion star and a neutron star \cite{Rawls:2011jw} with observational values of radius $\mathfrak{R} = 9.56_{-0.08}^{+0.08}$ km and mass $\mathcal{M} = 1.77 \pm 0.08~\mathcal{M}_{\odot}$. Sako et al. proposed an initial design for the Vela X-1's global X-ray emission line spectrum \cite{Sako:2000ve}. Since its discovery, it has received a great deal of attention among the compact objects in this field of study. Here we consider the mass and radius of this compact object to develop a new physically stable and realistic stellar model in this modified gravity by investigating several stellar properties.}\par
This paper is organized as follows: In Section \ref{sec2}, $f(R, T)$ gravity is defined mathematically in terms of isotropic matter distributions.
A conceivable $f(R,\, T)$ gravity model is displayed in Section \ref{sec3}. Section \ref{sec4} deals with boundary conditions; right here, we compare the interior and Schwarzschild's exterior metrics to determine the values of the unknown constant for the selected values of our model parameters. 
Section \ref{sec5} executes graphical analysis to investigate various physical properties and viability of compact stars.
The last section is devoted to the conclusions.
Throughout the investigation, we have mostly utilized a negative metric signature $(+, -, -, -)$, and \textcolor{blue}{followed the gravitational or geometrized unit system ($c = 1 = G$) throughout our manuscript except for the calculations in Table \ref{tb1}.}

\section{Interior Spacetime and Basic field Equations}\label{sec2}

Here, we focus on $f(R, T)$ gravity, a more comprehensive modified theory of gravity that incorporates geometry and matter coupling.
For $f(R,\,T)$ gravity, Harko {\em et al.} \cite{harko2011f} suggested the Hilbert-Einstein action term such as,

\begin{eqnarray}\label{action}
\mathcal{I} &=&\frac{1}{16 \pi}\int  f(R,T)\sqrt{-g} d^4 x + \int \mathcal{L}_m\sqrt{-g} d^4 x,
\end{eqnarray}
where $f ( R, T )$ represents arbitrary function of trace $T$ as well as Ricci scalar $R$, $g = det(g_{\mu \nu}$) and $\mathcal{L}_m$ being the lagrangian matter density which suggests the possibility of a non-minimal coupling between matter and geometry. The stress-energy tensor of the matter ($T_{\mu\nu}$) can be obtained from the matter
Lagrangian density $\mathcal{L}_m$ as \cite{landau2013classical},
\begin{eqnarray}\label{tmu1}
T_{\mu \nu}&=&-\frac{2}{\sqrt{-g}}\frac{\delta \sqrt{-g}\mathcal{L}_m}{\delta \sqrt{g^{\mu \nu}}}.
\end{eqnarray}
Trace of the matrix is given by $T=g^{\mu \nu}T_{\mu \nu}$. This can be further simplified by assuming that $\mathcal{L}_m$ does not depend on its derivatives but solely on $g_{\mu \nu}$. So, eqn.(\ref{tmu1}) becomes,
\begin{eqnarray}
T_{\mu \nu}&=& g_{\mu \nu}\mathcal{L}_m-2\frac{\partial \mathcal{L}_m}{\partial g^{\mu \nu}}.
\end{eqnarray}
\textcolor{blue}{Now, the variation of the Ricci scalar provides the following expression:}
\begin{eqnarray}
\delta R &=& R_{\mu \nu}\delta g^{\mu \nu} +  g_{\mu \nu}\Box \delta g^{\mu \nu} - \nabla_{\mu}\nabla_{\nu}\delta g^{\mu \nu},
\end{eqnarray}
and the variation of the action term (\ref{action}) can be written as:
\begin{eqnarray}\label{varI}
\delta \mathcal{I} &=& \frac{1}{16\pi}\int[f_R( R_{\mu \nu}\delta g^{\mu \nu} +  g_{\mu \nu}\Box \delta g^{\mu \nu} - \nabla_{\mu}\nabla_{\nu}\delta g^{\mu \nu}) + f_T\frac{\delta(g^{\alpha\beta}T_{\alpha\beta})}{\delta g^{\mu\nu}} - \frac{1}{2}g_{\mu\nu}f(R, T)\delta g^{\mu\nu} \nonumber\\&&
+ \frac{16\pi}{\sqrt{-g}}\frac{\delta(\sqrt{-g}\mathcal{L}_m)}{\delta g^{\mu\nu}}]\sqrt{-g}d^4 x,   
\end{eqnarray}
where, $f_R=\frac{\partial f(R,T)}{\partial R},~f_T=\frac{\partial f(R,T)}{\partial T}$.\\
$\nabla_{\nu}$ represents the covariant derivative
associated with the Levi-Civita connection of $g_{\mu \nu}$, $\Theta_{\mu \nu}=g^{\alpha \beta}\frac{\delta T_{\alpha \beta}}{\delta g^{\mu \nu}}$ and
$\Box \equiv \frac{1}{\sqrt{-g}}\partial_{\mu}(\sqrt{-g}g^{\mu \nu}\partial_{\nu})$ represents the D'Alembert operator.\\

The field equations of $f(R, T)$ gravity theory are now obtained by performing by parts integration to the second and third terms in the r.h.s. of Equation (\ref{varI}) as,
\begin{eqnarray}\label{frt}
f_R R_{\mu \nu}-\frac{1}{2}f(R,T)g_{\mu \nu}+(g_{\mu \nu }\Box-\nabla_{\mu}\nabla_{\nu})f_R &=&8\pi T_{\mu \nu}-f_T T_{\mu \nu}-f_T \Theta_{\mu \nu}
\end{eqnarray}
It is now possible to acquire the divergence of the stress-energy tensor $T_{\mu \nu}$ by taking the covariant divergence of (\ref{frt})\cite{harko2011f,koivisto2006note} as
\begin{eqnarray}\label{conservation}
\nabla^{\mu}T_{\mu \nu}&=&\frac{f_T}{8\pi-f_T}\left[(T_{\mu \nu}+\Theta_{\mu \nu})\nabla^{\mu}\ln f_T+\nabla^{\mu}\Theta_{\mu \nu} - \frac{1}{2}g_{\mu\nu}\nabla^{\mu}T\right].
\end{eqnarray}
\textcolor{blue}{We can infer from the above equation that there is a chance that the matter energy-momentum tensor will not be conserved under this modified theory, in contrast to GR. An additional force is produced as a result of the geometry-matter coupling \cite{koivisto2006note}. This is a common characteristic of the $f(R, T)$ theories. This matter non-conservation may involve quantum effects and particle creation in cosmological scenario. This physical interpretation makes the most practical sense. However, the possibility of active matter-creation processes in the universe remains a hypothesis without any direct observational or experimental evidence. For a detailed discussion of this aspect in the context of $f(R, T)$ gravity, see Ref. \cite{Lobato:2018vpq}. This non-conservation may occur when extending flat space expressions to curved space, resulting in a non-trivial conservation law. It is also possible to construct a conservative $f(R, T)$ gravity by making specific choices in the $f(T)$ counterpart, as illustrated in Ref. \cite{dos2019conservative, Pretel:2021kgl, Carvalho:2019gzs}.
From Eqn.(\ref{conservation}), we can easily check that the covariant derivative of the
stress-energy tensor i.e. $\nabla^{\mu}T_{\mu \nu}\neq 0$ if $f_T(R,T)\neq 0.$ For this present model, $f_T(R, T) = 2\chi$ and hence $\nabla^{\mu}T_{\mu \nu}\neq 0$.  Hence, the system will not be conserved for our chosen $f(R,\,T)$ model.}
Moreover, when $f(R, T)=f(R)$, from eqn. (\ref{frt}) we derive the $f(R)$ gravity field equations.\par
We consider the energy-momentum tensor in this study to be that of a perfect fluid i.e.
\begin{eqnarray}
T_{\mu \nu}&=&(p+\rho)u_{\mu}u_{\nu}-p g_{\mu \nu},
\end{eqnarray}
where $p$ denotes isotropic pressure and $\rho$ denotes matter density under modified gravity. The fluid velocity, $u^{\mu}$, fulfils the following equations: $u^{\mu}u_{\mu}=1$ and $u^{\mu}\nabla_{\nu}u_{\mu}=0$. As proposed by Harko et al. \cite{harko2011f}, we have taken the matter Lagrangian that can be taken as $\mathcal{L}_m=-p$ and the expression of $\Theta_{\mu \nu}=-2T_{\mu \nu}-pg_{\mu\nu}.$\\
Here, we take a look at a static, spherically symmetric line element in curvature coordinates $(t,r,\theta,\phi)$ provided to comprehend the internal geometry of compact star objects such as: 
\begin{equation}\label{line}
ds^{2}=-e^{\nu (r)}dt^{2}+e^{\lambda (r)}dr^{2}+r^{2}d\Omega^{2},
\end{equation}
where $d\Omega^{2}\equiv \sin^{2}\theta d\phi^{2}+d\theta^{^2}$. The metric coefficients $\nu(r)$ and $\lambda(r)$ are purely radial functions.

In this paper, we assume the form of $f(R, T )$ as we choose
\begin{eqnarray}\label{e}
f(R,T)&=& R+2 \chi T,
\end{eqnarray}
where $\chi$ is the coupling parameter, which varies based on the physical characteristics of our current model.\\
Using (\ref{e}) into (\ref{frt}),
the field equations in $f(R,T)$ gravity is given by,
\begin{eqnarray}
G_{\mu \nu}&=& 8\pi T_{\mu \nu}+ \chi T g_{\mu \nu} + 2\chi (T_{\mu \nu}+p g_{\mu \nu}),
\end{eqnarray}
where $G_{\mu \nu}$ is the Einstein tensor. 
With the help of equation(\ref{metric}), we can find the non-zero components of the Einstein
tensors $G^0_0$, $G^1_1$, $G^2_2 = G^3_3$ and finally get the 
field equations in modified gravity as,
\begin{eqnarray}
G^0_0: 8\pi\rho^{\text{E}}&=&\frac{\lambda'}{r}e^{-\lambda}+\frac{1}{r^{2}}(1-e^{-\lambda}),\label{f1}\\
G^1_1: 8 \pi p^{\text{E}}&=& \frac{1}{r^{2}}(e^{-\lambda}-1)+\frac{\nu'}{r}e^{-\lambda},\label{f2} \\
G^2_2 = G^3_3: 8 \pi p^{\text{E}}&=&\frac{1}{4}e^{-\lambda}\left[2\nu''+\nu'^2-\lambda'\nu'+\frac{2}{r}(\nu'-\lambda')\right]. \label{f3}
\end{eqnarray}
 
Here $\rho^{\text{E}}$ and $p^{\text{E}}$ are respectively the density and pressure in Einstein Gravity and
\begin{eqnarray}
\rho^{\text{E}}&=& \rho+\frac{\chi}{8\pi}(3 \rho-p),\label{r1}\\
p^{\text{E}}&=& p-\frac{\chi}{8\pi}(\rho-3p),\label{r2}
\end{eqnarray}
the prime ($'$) indicates differentiation with respect to `r'. Using Eqs. (\ref{f1})-(\ref{f3}), we get,
\begin{eqnarray}\label{con}
\frac{\nu'}{2}(\rho+p)+\frac{dp}{dr}&=&\frac{\chi}{2\chi + 8\pi}(p'-\rho').
\end{eqnarray}
We obtain the Einstein gravity conservation equation for $\chi = 0$ in equation(\ref{con}).

\section{Exact Solution of our proposed Model for isotropic Stars}\label{sec3}

There are numerous methods available in the literature for characterizing the inner area of a compact star structure. Also, our objective is to solve the system of equations (\ref{f1})-(\ref{f3}) to obtain a model of a compact star. The well-known Heintzmann IIa {\em ansatz} \cite{heintz, Delgaty:1998uy} is utilized, which includes nearly all of the solutions known for the static Einstein equations with a perfect fluid source, as provided by,
\begin{eqnarray}\label{elambda}
e^{\lambda}&=&  \frac{1}{\Bigg[1- \frac{3 B  r^2}{2} \Bigg\{\frac{1+\frac{C}{\sqrt{1+4 B r^2}}}{1 + B r^2}\Bigg\} \Bigg]}            ,
\end{eqnarray}
and,
\begin{eqnarray}\label{enu}
e^{\nu}&=& A^2 (1 + B r^2)^3,
\end{eqnarray}
in which the constants $A$ and $C$ are dimensionless and $B$ is with a dimension of $length^{-2}$. From the matching conditions, they can be evaluated. \\
Utilizing the following metric potentials, one can compute the Einstein density and pressure as:
\begin{eqnarray}
\rho^{\text{E}}&=&\frac{3 B \bigg \{(3 + B r^2) (1 + 4 B r^2)^{3/2} + C (3 + 9 B r^2)\bigg \}}{16 \pi (1 + B r^2)^2 (1 + 4 B r^2)^{3/2}},\\
p^{\text{E}}&=& -\frac{ 3 B  \bigg\{C + 7 B C r^2 + 3 (-1 + B r^2) \sqrt{1 + 4 B r^2} \bigg\}}{16 \pi (1 + B r^2)^2 \sqrt{1 + 4 B r^2}},
\end{eqnarray}
The physical parameters, like as Einstein pressure and density, must be obtained before so that we can investigate the full structure of stellar models.

By utilizing the $p^{\text{E}}$ and $\rho^{\text{E}}$ expressions from equations (\ref{r1}) and (\ref{r2}), we can derive the matter density and pressure expressions $\rho,\,p$ in modified gravity as,
\begin{eqnarray}
\rho&=& \frac{ 3 B \Big[(1 + 4 B r^2)^{3/2} \Big(3 \chi + 2\pi (3 + B r^2)\Big) + 
   C \Big\{6 (\pi + 3 B \pi r^2) + \chi \Big(2 + B r^2 (4 - 7 B r^2)\Big)\Big\}\Big] }{4 (\chi + 2 \pi) (\chi + 4 \pi) (1 + B r^2)^2 (1 + 4 B r^2)^{3/2}},\label{p1}\\
      p&=& -\frac{ 3 B \Big[(1 + 4 B r^2)^{3/2} \Big(-3 (\chi + 2 \pi) + 2 B (\chi + 3 \pi) r^2\Big) + 
   C \Big\{2 \pi + 2 B (3 \chi + 11 \pi) r^2 + 7 B^2 (3 \chi + 8 \pi) r^4\Big\}\Big] }{4 (\chi + 2 \pi) (\chi + 4 \pi) (1 + B r^2)^2 (1 + 4 B r^2)^{3/2}}. \label{p2}
\end{eqnarray}

\section{Exterior line element and matching conditions}\label{sec4}

In addition to the fact that there should be no pressure on the surface of a celestial structure, fundamental junction conditions ensure harmonious matching at the boundary for the exterior and interior solutions of a static stellar object. The Schwarzschild exterior solution at the boundary, or $r= \mathfrak{R}$ (Radius of the star), will match our internal spacetime at $r=\mathfrak{R}$ of the star. In addition, the asymptotically flat aspect of the Schwarzschild vacuum solution makes it significant for astrophysics.
Corresponding to the interior spacetime,
 \begin{eqnarray}
ds_{-}^2&=& -A^2 (1 + B r^2)^3 dt^2+ \frac{1}{\Bigg[1- \frac{3 B  r^2}{2} \Bigg\{\frac{1+\frac{C}{\sqrt{1+4 B r^2}}}{1 + B r^2}\Bigg\} \Bigg]}  dr^2  + ~ r^2(d\theta^2 +\sin^2 \theta d\phi^2),
\end{eqnarray}
the exterior line element is stated as follows:
\begin{eqnarray}
ds_+^{2}&=&-\left(1-\frac{2\mathcal{M}}{r}\right)dt^{2}+\left(1-\frac{2\mathcal{M}}{r}\right)^{-1}dr^{2}+r^{2}\left(d\theta^{2}+\sin^{2}\theta d\phi^{2}\right),
\end{eqnarray}
The term '$\mathcal{M}$' indicates the entire mass within the boundary of a compact star.\\
Since the metric potentials at the boundary surface ($r= \mathfrak{R}$) of the strange star are continuous, the following formulations arise such as:
$$g_{rr}^+=g_{rr}^-,\, ~~\text{and} ~~~~g_{tt}^+=g_{tt}^-,$$
where ($-$) and ($+$) sign represent interior and exterior spacetime, respectively. The above two relationships imply,
\begin{eqnarray}
\left(1-\frac{2\mathcal{M}}{\mathfrak{R}}\right)^{-1}&=&\frac{1}{\Bigg[1- \frac{3 B  \mathfrak{R}^2}{2} \Bigg\{\frac{1+\frac{C}{\sqrt{1+4 B \mathfrak{R}^2}}}{1 + B \mathfrak{R}^2}\Bigg\} \Bigg]} ,\label{o1}\\
1-\frac{2\mathcal{M}}{\mathfrak{R}}&=&  A^2 (1 + B \mathfrak{R}^2)^3,\label{o2}
\end{eqnarray}
It is also required that the isotropic pressure `$p$' vanishes at the boundary $\mathfrak{R}$ i.e., $p(r=\mathfrak{R})=0$, implies the following equation:
\begin{eqnarray}
(1 + 4 B \mathfrak{R}^2)^{3/2} \Big(-3 (\chi + 2 \pi) + 2 B (\chi + 3 \pi) \mathfrak{R}^2\Big) + C \Big(2 \pi + 2 B (3 \chi + 11 \pi) \mathfrak{R}^2 + 
      7 B^2 (3 \chi + 8 \pi) \mathfrak{R}^4\Big) &=& 0 .\label{o3}
    \end{eqnarray}
 We obtain the expressions for $A$ and $C$ as a function of $B$ by solving the equations (\ref{o1})-(\ref{o3}) simultaneously,
  \begin{eqnarray}
 A &=& \sqrt{\frac{-2 \mathcal{M} + \mathfrak{R}}{\mathfrak{R} (1 + B \mathfrak{R}^2)^3}}~~[\textcolor{blue}{\text{Considering only positive signed expression}}],\label{aa}\\
       C&=& \frac{(1 + 4 B \mathfrak{R}^2)^{3/2} \Big(3 \chi + 6 \pi - 2 B (\chi + 3 \pi) \mathfrak{R}^2\Big)} {2 \pi + 2 B (3 \chi + 11 \pi) \mathfrak{R}^2 + 7 B^2 (3 \chi + 8 \pi) \mathfrak{R}^4} ,\label{cc}
      \end{eqnarray}
\textcolor{blue}{By fixing a preferred value of $B$ ($=0.001 ~\rm{km}^{-2}$), the values of $A$ and $C$ have been listed in Table~\ref{tb12} for various values of $\chi$, determined using the approximate mass and radius of the compact stellar candidate Vela X-1. From the expression of A in (\ref{aa}), we see that it is a function of $M, R$, and $B$. So, ultimately, A becomes a constant for this particular stellar model Vela X-1 after fixing a particular value of $B$. Thus $A$ is independent of $\chi$ throughout our model. On the contrary, the values of $C$ increase as $\chi$ increases.}

\begin{table*}[t]
\centering
\caption{The numerically derived values of the constants $A$ and $C$ for the compact star Vela X-1 for different values of coupling constant $\chi$(Taking $B=0.001 ~\rm{km}^{-2}$).}
\label{tb12}
\begin{tabular}{@{}|ccc|c|c|c|cccccccccc@{}}
\hline
Objects  & Estimated &Estimated & $\chi$ & $A$& $C$   \\
&Mass ($M_{\odot}$)& Radius &&&  \\
\hline\hline
Vela X-1  \cite{Rawls:2011jw}& $1.77$&$9.56$& 0.00& 0.590838&$1.94257$\\
&&& $0.25$&0.590838 &$1.99677$                                    \\
&&& $0.50$ &0.590838  &$2.04961$\\
&&& $0.75$ &0.590838  &$2.10115$\\
&&& $1.00$ &0.590838 &$2.15142$\\
&&& $1.25$ &0.590838 &$2.20048$\\
\hline
\end{tabular}
\end{table*}

\section{Physical properties of the astrophysical structure in $f(R,\,T)$ gravity theory}\label{sec5}

In this section, we will test physical highlights of the stellar structure in the context of $f(R,\, T)$ theory to investigate the Modified TOV equation, energy conditions, the status of the sound speed within the stellar system, compactness and gravitational surface redshift, the adiabatic index, and so on for different values of the coupling constant $\chi$.

\subsection{Regularity of the metric potentials}\label{mp}

For the model to be physically viable and stable, the metric potentials inside the compact stellar structure should be positive, monotonically increasing, singularity-free, and regular. So, to find out if there are singularities, we now study the behavior of both potentials within the range (0, $\mathfrak{R}$). 
\\
From equation (\ref{elambda}-\ref{enu}) we get $e^{\lambda}=1$ and $e^{\nu}=A^2$, at the center of the star. Also, their derivatives are given by,
  \begin{eqnarray}
  (e^{\lambda})'&=& \frac{12 B r \{C + 2 B C r^2 - 2 B^2 C r^4 + (1 + 4 B r^2)^\frac{3}{2}\}}{\sqrt{1 + 4 B r^2} \{-2\sqrt{1 + 4 B r^2} + 
   B r^2 (3 C + \sqrt{1 + 4 B r^2})\}^2},\\
  (e^{\nu})'&=& 6Br(A + ABr^2)^2,
 \end{eqnarray}
 \\
  The derivatives of the metric coefficients vanish at the center of the star, implying that the metric coefficients are regular at the center of the star. \begin{figure}[htbp]
    \centering
        \includegraphics[scale=.65]{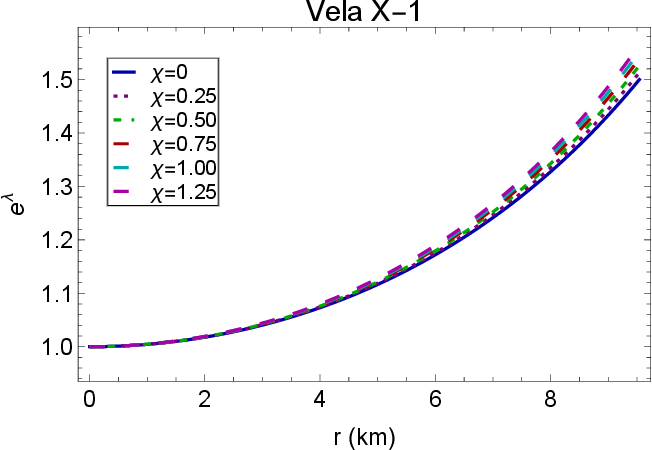}
        \includegraphics[scale=.65]{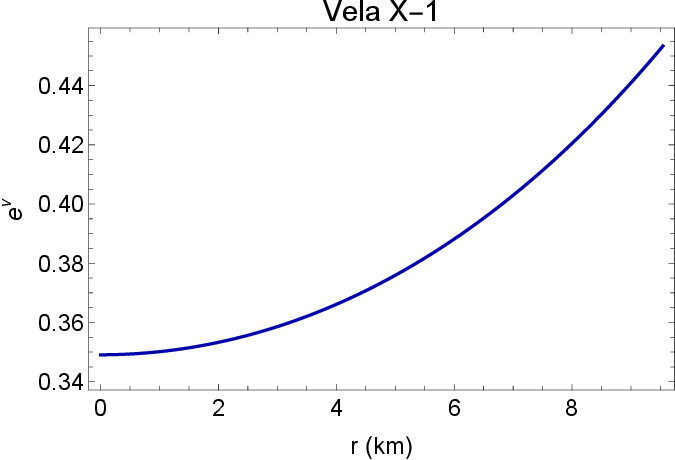}
       \caption{$e^{\lambda}$ and $e^{\nu}$ are shown against `r'.}\label{metric}
\end{figure}
It is found that both metric potentials are consistent with the previously indicated requirements. The properties of metric coefficients are illustrated in FIG.~\ref{metric}.
Both metric potentials have minimum value at the center and a nonlinear increase in value until they reach their maximum at the boundary of the surface, as shown by the graphical behavior.
\newpage

\subsection{Nature of pressure and density}
\begin{figure}[htbp]
    \centering
        \includegraphics[scale=.65]{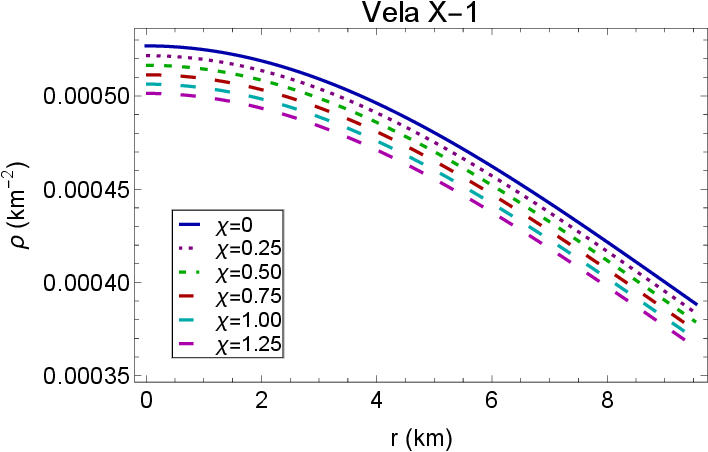}
        \includegraphics[scale=.65]{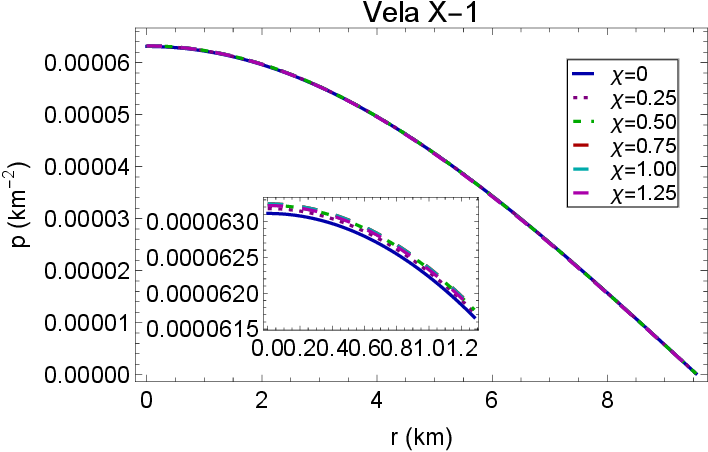}
        \caption{(left) Matter density and (right) isotropic pressure are plotted against radius for different values of the coupling constant mentioned in the figure. \label{pp}}
\end{figure}

For various values of $\chi$, the matter density and isotropic pressure profiles are shown in FIG.~\ref{pp}. These pictures further demonstrate the absence of physical and mathematical singularities in our framework by exhibiting positive and regular energy densities and pressures at the origin. It is evident that the two physical variables reach their maximum at the origin and then fall monotonically to their minimum values at the surface. It is also seen that at the boundary$(r= \mathfrak{R})$, pressure vanishes but density remains positive i.e. $p(r = \mathfrak{R}) = 0, \rho(r = \mathfrak{R}) \neq 0$. So ultimately, these indicate that the proposed stellar model is conceivable. 
\\
Also, we can calculate the central pressure and central density \textcolor{blue}{under modified gravity} as
\begin{eqnarray}
\rho_c &=& \rho\Big\rvert_{r=0} = \frac{3B}{4}\Bigg[\frac{C}{\chi + 2 \pi} + \frac{3+C}{\chi + 4\pi}\Bigg],\\
 p_c&=& p\Big\rvert_{r=0} =  \frac{3B\{3 \chi - 2 (-3 + C) \pi\}}{4(\chi + 2 \pi)(\chi + 4\pi)}.
\end{eqnarray}
Clearly, $\rho_c$ and $p_c$ are obviously finite.\\ 
\textcolor{blue}{ Now we are interested in evaluating the proper range of the coupling parameter $\chi$.\\ 
\begin{center}
  \bf{\underline{Case I}}  
\end{center}
We initially assume that, $\chi+2\pi>0$.\\
Now $p_c>0 \implies \chi > -\frac{2(3 - C)\pi}{3}$.\\
So we obtain,
\begin{equation}\label{1st}
\chi > Max\Bigg\{-2\pi, -\frac{2(3 - C)\pi}{3}\Bigg\}.    
\end{equation}
Also since, $\rho_c - p_c > 0 \implies \frac{3BC}{2\chi+4\pi} > 0 \implies C> 0.$\\
Again, $\rho_c > 0 \implies \chi > -\frac{6(1 + C)\pi}{2 + 3C}$. So, it is clear that $\Bigg[-\frac{2(3 - C)\pi}{3} - \Bigg\{-\frac{6(1 + C)\pi}{2 + 3C}\Bigg\}\Bigg] > 0$, for $C>0$. So, with the help of (\ref{1st}) and $C>0$, again we can repeat, 
$\chi > Max\Bigg\{-2\pi, -\frac{2(3 - C)\pi}{3}\Bigg\}$.\\ 
Also, $C>0 \implies -\frac{2(3 - C)\pi}{3} > -2\pi$.\\
Hence, we can finally obtain, 
\begin{equation}\label{11st}
\chi > -\frac{2(3 - C)\pi}{3}.
\end{equation}
\begin{center}
  \bf{\underline{Case II}}  
\end{center}
Let us assume, $\chi + 4\pi < 0 \implies \chi + 2\pi < 0$ also.\\
Now, $\rho_c - p_c > 0 \implies C < 0$.
Again, $p_c > 0 \implies \chi > -\frac{2(3 - C)\pi}{3}$.\\
So, with the help of $C < 0$ we obtain,
\begin{equation}\label{2nd}
-\frac{2(3 - C)\pi}{3} <\chi < -2\pi.
\end{equation}
\begin{center}
  \bf{\underline{Case III}}  
\end{center}
Now let us assume, $-4\pi < \chi < -2\pi$.\\
Now, $p_c >0 \implies \chi < -\frac{2(3 - C)\pi}{3}$.\\
Now, $\rho_c - p_c > 0 \implies C < 0.$ So, obviously $-\frac{2(3 - C)\pi}{3} < -2\pi$.\\
Hence, we obtain the range for $\chi$ as,
\begin{equation}
-4\pi < \chi < Min\Bigg\{ -2\pi, -\frac{2(3 - C)\pi}{3}\Bigg\}.
\end{equation}}
The expressions of $\rho$ and $p$ given in equations (\ref{p1} - \ref{p2}) can be differentiated to obtain the pressure and density gradient. This provides,
\begin{eqnarray}
\frac{d\rho}{dr}&=& -\frac{f_1(r)}{(\chi + 
       2 \pi) (\chi + 4 \pi) (1 + B r^2)^3 (1 + 4 B r^2)^\frac{5}{2}} ,\\
\frac{dp}{dr}&=& \frac{f_2(r)}{(\chi + 
     2 \pi) (\chi + 4 \pi) (1 + B r^2)^3 (1 + 4 B r^2)^\frac{5}{2}}.
\end{eqnarray}
where $$f_1(r) = 3 B^2 r \bigg[(1 + 4 B r^2)^\frac{5}{2} \big\{3 \chi + \pi (5 + B r^2)\big\} + 
       3 C \Big[\pi \big\{5 + B r^2 (23 + 30 B r^2)\big\} + 
          \chi \big[2 + B r^2 \big\{9 + B r^2 (9 - 7 B r^2)\big\}\big]\Big]\bigg]$$
          and $$f_2(r) = 3 B^2 r \bigg[-3 C (\chi + \pi) - 3 BC (4 \chi + 3 \pi) r^2 + 
     9 B^2 C (\chi + 6 \pi) r^4 + 
     21 B^3 C (3 \chi + 8 \pi) r^6 + (1 + 4 B r^2)^\frac{5}{2} \big\{-4 \chi - 9 \pi + B (\chi + 3 \pi) r^2\big\}\bigg].$$
\begin{figure}[htbp]
    \centering
        \includegraphics[scale=.65]{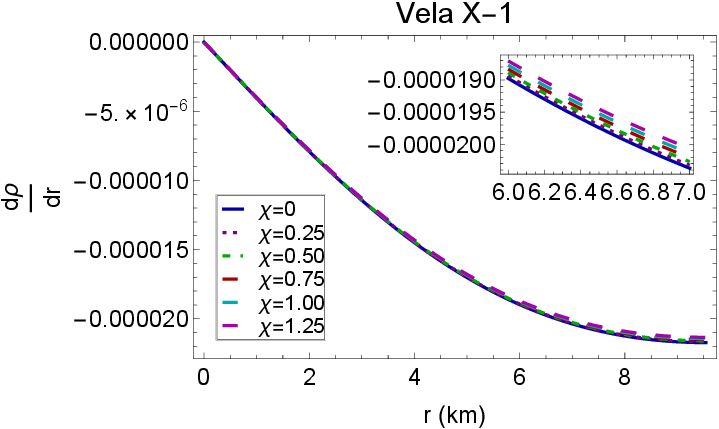}
        \includegraphics[scale=.65]{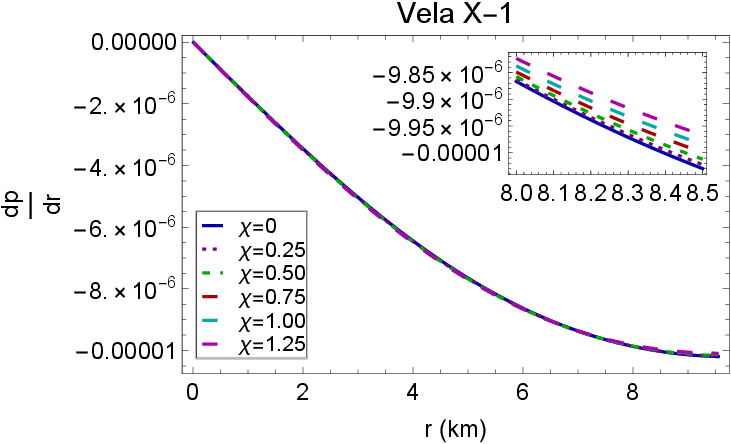}
        \caption{The density and pressure gradients ($\frac{d\rho}{dr}$ and  $\frac{dp}{dr}$) are plotted against `r'.}\label{grad5}
\end{figure}
\begin{figure}[htbp]
    \centering
        \includegraphics[scale=.65]{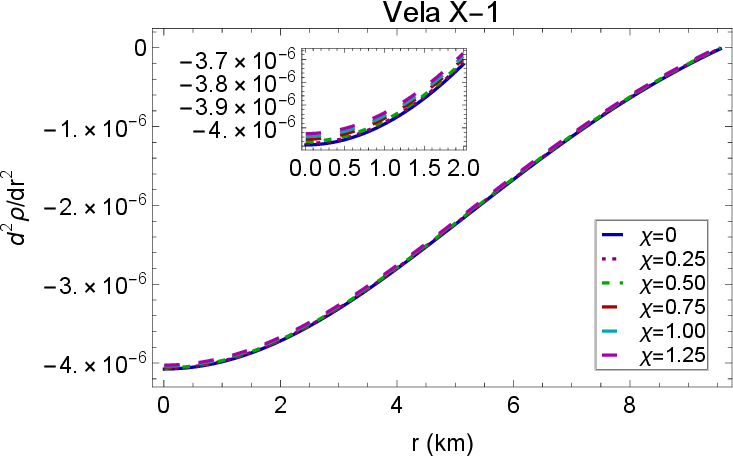}
        \includegraphics[scale=.65]{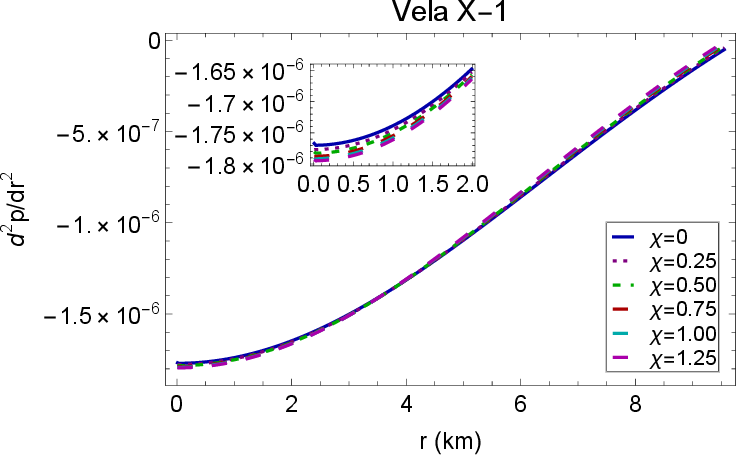}
       \caption{Second order derivatives of density and pressure ($\frac{d^2\rho}{dr^2}$ and $\frac{d^2 p}{dr^2}$) are plotted against `r'.}\label{rho2}
\end{figure}

FIG.~\ref{grad5} illustrates the behavior of the pressure and density gradient for various values of $\chi$. From the figure FIG.~\ref{grad5}, one can note that $\frac{d\rho}{dr},\frac{dp}{dr}<0$ which illustrates how density and pressure fall as the radius of the compact object increases. Furthermore, it has been shown in FIG.~\ref{rho2} that the central density and pressure can reach their maximum at $r = 0$, suggesting that $\frac{d\rho}{dr}=\frac{dp}{dr}=0$ and $\frac{d^2\rho}{dr^2}<0,\frac{d^2p}{dr^2}<0$.

\begin{table*}[t]
\centering
\caption{Numerically computed values of central density($\rho_c$), surface density($\rho_s$), central pressure($p_c$) for the compact star Vela X-1 for different values of coupling constant $\chi$ (Taking $B=0.001 ~\rm{km}^{-2}$).}
\label{tb1}
\begin{tabular}{@{}|c|c|c|c|@{}}
\hline
$\chi$& $\rho_c $ ($\text{gm}~\text{cm}^{-3}$) & $\rho_s$ ($\text{gm}~\text{cm}^{-3}$) & $p_c$ ($\text{dyne}~\text{cm}^{-2}$) \\
\hline\hline
 0.00& $7.10911 \times 10^{14}$& $5.23607 \times 10^{14}$ & $7.66415 \times 10^{34}$\\
 0.25& $7.03851 \times 10^{14}$ & $5.16959 \times 10^{14}$& $7.67245 \times 10^{34}$\\
 0.50& $6.96879 \times 10^{14}$ & $5.10477 \times 10^{14}$ & $7.6778 \times 10^{34}$\\
 0.75& $6.89999 \times 10^{14}$ & $5.04156 \times 10^{14}$ & $7.68033 \times 10^{34}$\\
 1.00& $6.83214 \times 10^{14}$ & $4.97989 \times 10^{14}$ & $7.68019 \times 10^{34}$\\
 1.25& $6.76524 \times 10^{14}$& $4.91972 \times 10^{14}$ & $7.67752 \times 10^{34}$\\
\hline
\end{tabular}
\end{table*}

\newpage

 \subsection{Energy Conditions}
 \textcolor{blue}{Pointwise energy conditions were first proposed as physically plausible constraints on matter within the framework of mathematical relativity. For example, they seek to convey the attractiveness of gravity or the positivity of mass. More significantly, they have served as presumptions in mathematical relativity to support the non-existence of wormholes and other exotic phenomena as well as singularity theorems. Energy conditions play a significant impact in the context of the exotic properties. These conditions restrict the contraction of the stress tensor at each spacetime point. The null energy condition (NEC), the weak energy condition (WEC), the strong energy condition (SEC), and the dominant energy condition (DEC) are characterized as the four main energy restrictions and are crucial to investigate the existence of a realistic matter distribution. To rewrite these energy conditions in Einstein's equation, we can replace the stress tensor with the Ricci curvature tensor, resulting in a geometric form rather than the original physical form. Depending on the energy-momentum tensor $T_{\mu\nu}$, the physical forms of these Energy conditions are described as:}
\begin{itemize}
\item Strong Energy Condition (SEC):~ $(T_{\mu\nu} - \frac{1}{2}Tg_{\mu\nu})U^\mu U^\nu \geq 0$ i.e.~  $\rho+p \geq 0, ~\rho+ 3p \geq 0,$\\

\item Weak Energy Condition (WEC):~ $T_{\mu\nu}U^\mu U^\nu \geq 0$ i.e.~ $\rho+p\geq 0,~ \rho \geq 0,$\\

\item Null Energy Condition (NEC):~ $T_{\mu\nu}k^\mu k^\nu \geq 0$ i.e.~ $\rho+p\geq 0,$ \\

\item Dominant Energy Condition (DEC):~$T_{\mu\nu}U^\mu U^\nu \geq 0$ i.e.~ $\rho-p\geq 0,~ \rho \geq 0$. \\ 

\end{itemize}
Here, $k^\nu$ is a null vector and $U^\mu$ is a time-like vector. In this connection, it is to be mentioned that where $U^\mu$ is a time-like vector
but $T_{\mu\nu}U^\mu$ is not space-like.

So, we will need the following expressions to validate the energy conditions given earlier.

\begin{eqnarray}
\rho+p&=&\frac{3 B \Big[-C (-1 + B r^2 + 14 B^2 r^4) + 
   \sqrt{1 + 4 B r^2} \Big\{3 + B r^2 (11 - 4 B r^2)\Big\}\Big]}{2 (\chi + 
   4 \pi) (1 + B r^2)^2 (1 + 4 B r^2)^\frac{3}{2}},\\
\rho+3p&=&-\frac{3 B \Big[(1 + 4 B r^2)^\frac{3}{2}\Big\{-6 (\chi + 2 \pi) + B (3 \chi + 8 \pi) r^2\Big\} + 
       C \Big\{-\chi + B (7 \chi + 24 \pi) r^2 + 
          7 B^2 (5 \chi + 12 \pi) r^4\Big\}\Big]}{2 (\chi + 
   4 \pi) (1 + B r^2)^2 (1 + 4 B r^2)^\frac{3}{2}},\\
   \rho-p&=&\frac{3 B \Big\{C + B r^2 (1 + 4 B r^2)^\frac{3}{2} + 
   BC r^2 (5 + 7 B r^2)\Big\}}{2 (\chi + 2 \pi) (1 + B r^2)^2 (1 + 
   4 B r^2)^\frac{3}{2}}.
\end{eqnarray}

 \begin{figure}[htbp]
    \centering
        \includegraphics[scale=.46]{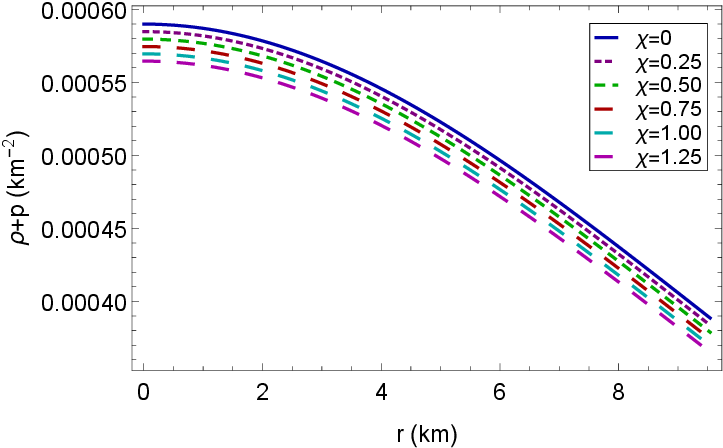}
        \includegraphics[scale=.45]{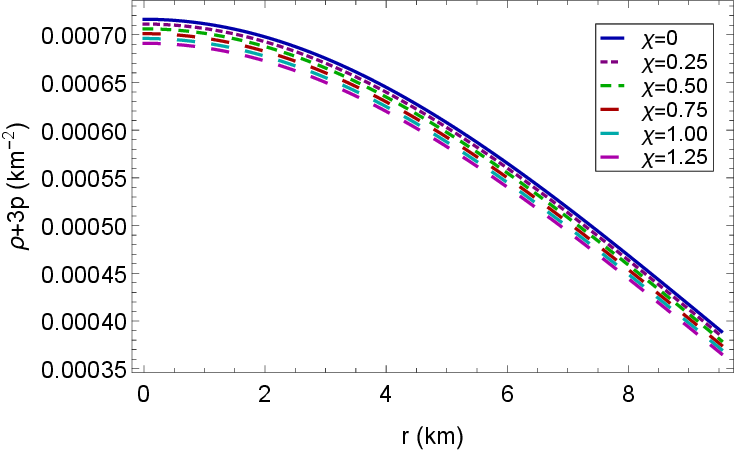}
        \includegraphics[scale=.45]{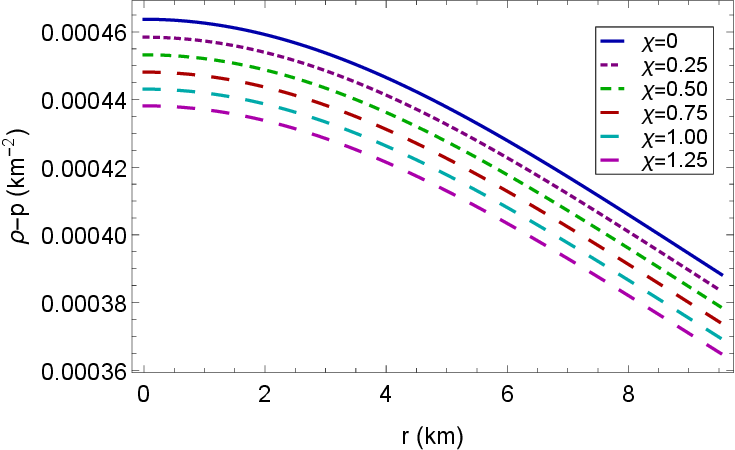}
       \caption{The energy conditions are shown against `r'.}\label{ener1}
\end{figure}
Our suggested $f(R, T)$ model is physically viable for all indicated values of the model parameter, as demonstrated graphically in FIG.~\ref{ener1} by the consistency of these energy bounds.

\subsection{Causality condition}

The stability of a physical structure under external oscillations can be used to assess its consistency. Checking the causality requirement allows us to examine the stability of our candidates for isotropic compact stars. To meet the causality criterion, the square of the sound speed $V^2$ throughout the fluid sphere has to correspond to the limitation $0<V^2<1$. Here
\begin{eqnarray}
V^2&=& \frac{dp}{d\rho} = \frac{f_3(r)}{f_4(r)},
\end{eqnarray}
where,
\begin{eqnarray*}
f_3(r) &=& -\Big[-3 C (\chi + \pi) - 3 B C (4 \chi + 3 \pi) r^2 + 
    9 B^2 C (\chi + 6 \pi) r^4 + 
    21 B^3 C (3 \chi + 8 \pi) r^6 + (1 + 4 B r^2)^\frac{5}{2} \Big\{-4 \chi - 9 \pi \\&& + B (\chi + 3 \pi) r^2\Big\}\Big]  ,\\
f_4(r) &=& (1 + 4 B r^2)^\frac{5}{2} \Big\{3 \chi + \pi (5 + B r^2)\Big\} + 
  3 C \Bigg[\pi \Big\{5 + B r^2 (23 + 30 B r^2)\Big\} + 
     \chi \Big[2 + B r^2 \Big\{9 + B r^2 (9 - 7 B r^2)\Big\}\Big]\Bigg].
\end{eqnarray*}

\begin{figure}[htbp]
    \centering
        \includegraphics[scale=.65]{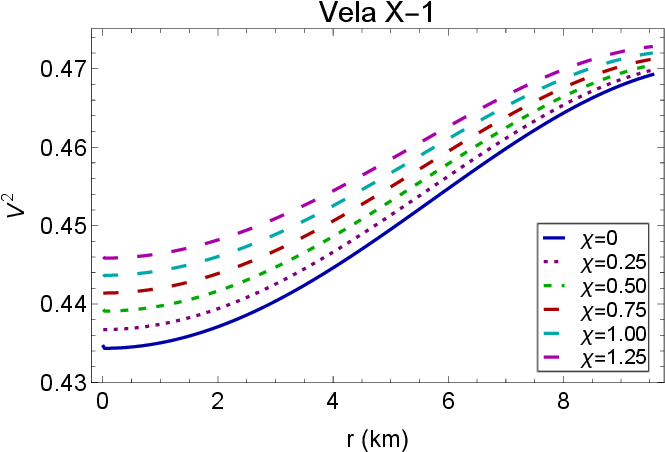}
        \includegraphics[scale=.65]{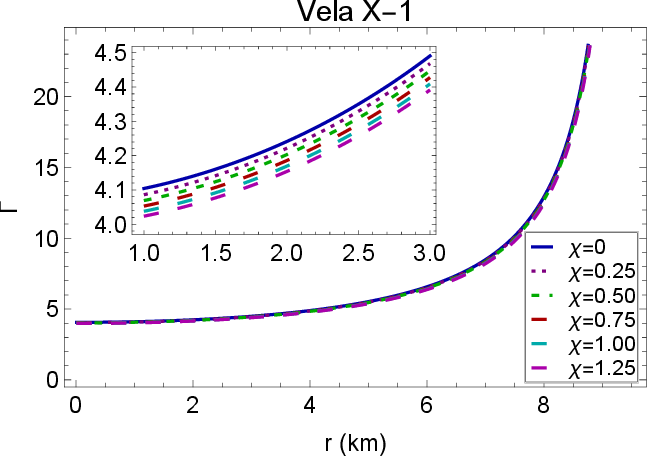}
       \caption{(left) The square of the sound velocity and (right) relativistic adiabatic index are plotted against the radius inside the stellar interior.\label{sv}}
\end{figure}
For this suggested stellar structure, we depict the evolution of the sound speed for different values of $\chi$ and find that the condition specified above has been fulfilled, as FIG. \ref{sv} illustrates. Our suggested compact star model is hence stable.

\subsection{Relativistic adiabatic index}
The adiabatic index, which is used to describe the stiffness of the EoS for a given energy density, is crucial when talking about the stability of both relativistic and non-relativistic compact objects \cite{sharifa2018anisotropic}.
\textcolor{blue}{ In order to derive the equations governing the radial oscillations both spacetime and fluid variables are perturbed so as not to violate the spherical symmetry of the background body. In the unperturbed model, if we define $\delta r(r,t)$ as the time-dependent radial displacement of a fluid element positioned at location $r$ and assuming a harmonic time dependence-
$$\delta r(r,t) = X(r)e^{i\omega t}$$ ($\omega$ be the frequency of the oscillation),
we derive the subsequent formula that characterizes the radial oscillations as in Ref. \cite{kokkotas2001radial}: 
$$V^2 X''+[(V^2)'- Z +4\pi r\gamma p e^{2\lambda} - \nu']X'+[2(\nu')^2 + \frac{2m}{r^3} - Z'- 4\pi(\rho + p)Z r e^{2\lambda} + \omega^2 e^{2\lambda -2\nu}]X=0.$$
Where $V$ is the sound speed, which is calculated from the unperturbed background for a specific EoS, 
$$V^2=\frac{dp}{d\rho}$$
and $\Gamma$ is the adiabatic index, which is connected to the sound speed through \begin{eqnarray}
\Gamma &=&\frac{\rho+p}{p}\frac{dp}{d\rho},\nonumber\\&=&\frac{f_5(r)}{f_6(r)}
\end{eqnarray}
where,
\begin{eqnarray*}
f_5(r) &=& -2 (\chi + 2 \pi) \Big[-3 C (\chi + \pi) - 
      3 B C (4 \chi + 3 \pi) r^2 + 9 B^2 C (\chi + 6 \pi) r^4 + 
      21 B^3 C (3 \chi + 8 \pi) r^6 \\&& + (1 + 4 B r^2)^\frac{5}{2} \big\{-4 \chi - 9 \pi + B (\chi + 3 \pi) r^2\big\}\Big] \Big[(-3 + 
         B r^2) (1 + 4 B r^2)^\frac{3}{2} + C (-1 + B r^2 + 14 B^2 r^4)\Big],\\
f_6(r) &=& \Big[(1 + 4 B r^2)^\frac{3}{2}\big\{-3 (\chi + 2 \pi) + 2 B (\chi + 3 \pi) r^2\big\} +
     C\big\{2 \pi + 2 B (3 \chi + 11 \pi) r^2 + 
       7 B^2 (3 \chi + 8 \pi) r^4\big\}\Big] \\&& \times\Big[(1 + 4 B r^2)^\frac{5}{2} \big\{3 \chi + \pi (5 + B r^2)\big\} + 
    3 C \big[\pi \big\{5 + B r^2 (23 + 30 B r^2)\big\} + 
       \chi [2 + B r^2 \{9 + B r^2 (9 - 7 B r^2)\}]\big]\Big].
\end{eqnarray*}
}
 According to the assessments of some researchers, within a dynamically stable stellar object, the adiabatic index value is expected to be greater than $\frac{4}{3}$ \cite{Chandrasekhar:1964zz, heintzmann1975neutron,hillebrandt1976anisotropic}.

The adiabatic index $\Gamma$ is illustrated in FIG.~\ref{sv}.
This concept indicates that even with greater curvature terms present in the $f(R, T)$ functional form, our considered isotropic stellar model stays within the stability range.

\subsection{Modified TOV Equation}

\textcolor{purple}{Every celestial object remains in static equilibrium, or dynamical balance, under different internal forces acting simultaneously on it. This condition is characterized by the generalized Tolman-Oppenheimer-Volkoff (TOV) equation. For our current isotropic stellar model in the $f(R, T)$ gravity system, the equilibrium condition is described by the following modified TOV equation.}
\begin{eqnarray}\label{con1}
-\frac{\nu'}{2}(\rho+p)-\frac{dp}{dr}+\frac{\chi}{2\chi + 8\pi}\Big(\frac{dp}{dr}-\frac{d\rho}{dr}\Big)=0,
\end{eqnarray}
From the above equation (\ref{con1}), the equilibrium condition under the combined behavior of various forces namely gravitational ($F_g$), hydrostatic ($F_h$), and the additional force due to the modified $f(R,\,T)$ gravity ($F_m$) for our compact star candidate can be checked. 
Equation (\ref{con1}) can be written as
\begin{eqnarray}\label{tov1}
F_g + F_h + F_m = 0
\end{eqnarray}
where,
\begin{eqnarray}
F_g &=& \frac{9 B^2 r \Big\{(-3 + B r^2) (1 + 4 B r^2)^\frac{3}{2} + 
   C (-1 + B r^2 + 14 B^2 r^4)\Big\}}{2 (\chi + 4 \pi) (1 + 
   B r^2)^3 (1 + 4 B r^2)^\frac{3}{2}},\\
F_h &=& \frac{3 B^2 r}{(\chi + 2 \pi) (\chi + 
     4 \pi) (1 + B r^2)^3 (1 + 4 B r^2)^\frac{5}{2}}\Bigg[(1 + 4 B r^2)^\frac{5}{2} \Big\{4 \chi + 9 \pi - B (\chi + 3 \pi) r^2\} + 
     3 C \{\chi + \pi \nonumber \\&&  + B (4 \chi + 3 \pi) r^2 - 
        3 B^2 (\chi + 6 \pi) r^4 - 
        7 B^3 (3 \chi + 8 \pi) r^6 \Big\}\Bigg],\\
   F_m &=& \frac{3 B^2 \chi r \Big[(-1 + B r^2) (1 + 4 B r^2)^\frac{5}{2} + 
     3 C \big[1 + B r^2 \big\{5 + 2 B r^2 (6 + 7 B r^2) \big\}\big]\Big]}{2 (\chi + 
     2 \pi) (\chi + 4 \pi) (1 + B r^2)^3 (1 + 4 B r^2)^\frac{5}{2}}.
   \end{eqnarray}
   
\begin{figure}[htbp]
    \centering
        \includegraphics[scale=.65]{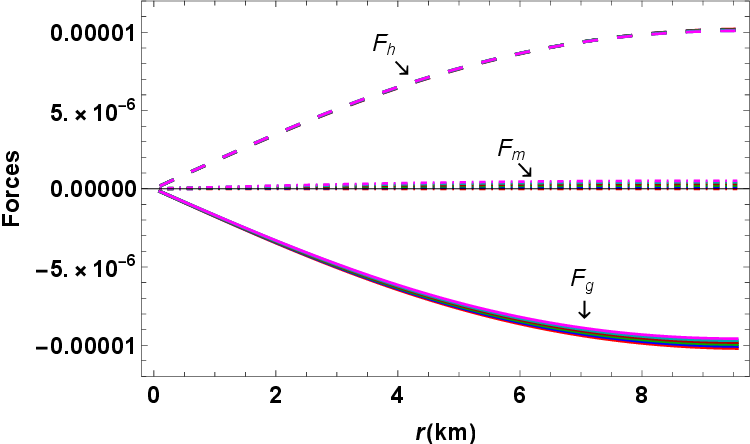}
       \caption{Different forces acting on the system are plotted against the radius inside the stellar interior for different values of $\chi$.  \label{tov1}}
\end{figure}

\textcolor{purple}{Ultimately, FIG.~\ref{tov1} evidently allows us to conclude that our present model attains its static equilibrium under these three different forces because the gravitational force counterbalances (or nullifies) the combined effect of hydrostatic force and modified gravity force.}\par
\textcolor{purple}{Although, the static equilibrium in consideration may exhibit stability or instability in response to a minor radial perturbation. Therefore, in the last subsection~(\ref{stable}), we have conducted the Harrison-Zeldovich-Novikov stability test in order to further investigate whether this static equilibrium is stable or unstable.}

\subsection{Equation of state (EoS)} 

An essential astrophysical tool that could help us comprehend the basic principles of matter dispersion is the EoS parameter. The EoS illustrates how matter is influenced by a suggested set of physical specifications. The relationship between pressure and matter density is simply explained by the equation of state. While several researchers used linear, quadratic, polytropic, and other EoS to represent the compact object. However, we did not make any particular assumptions about EoS in this work. Consequently, $\omega = p/\rho$ is denoted as the EoS parameter, which is commonly represented as a dimensionless number that may be used to characterize the relationship between matter density and pressure.
\begin{figure}[htbp]
    \centering
        \includegraphics[scale=.65]{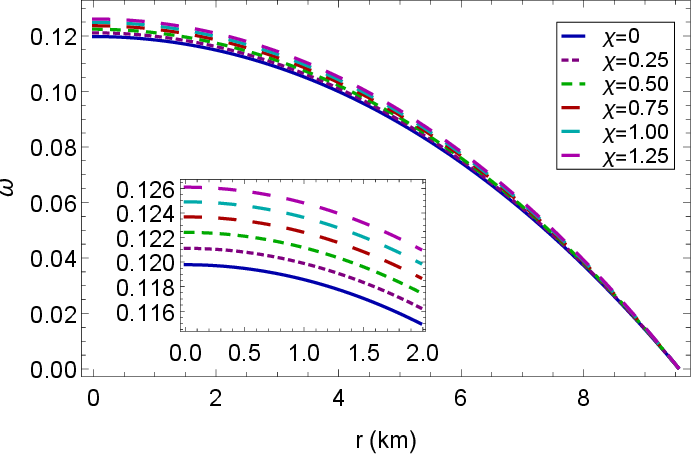}
        \includegraphics[scale=.68]{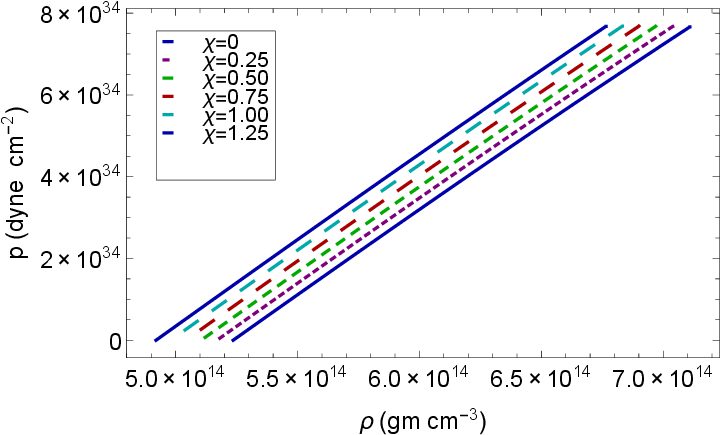}
       \caption{The pressure and density relation are shown inside the stellar interior.}\label{eoss}
\end{figure}\\
Its different values correspond to different stages of the entire cosmos. \textcolor{blue}{Regarding the stability of compact stellar objects, the Zeldovich requirement for stability is generally valid for a wide range of gravity theories, including modified ones. Several investigations use the Zeldovich condition for stability in modified gravity theories \cite{bhar2019compact}. Even though it is helpful, modified gravity theories cannot be guaranteed to be stable by the Zeldovich criterion alone. It gives a fundamental criterion based on gravitational and thermal energies but ignores additional aspects such as magnetic fields, rotation, and the detailed dynamics of gravity in modified theories. The gravitational potential and equations of motion in modified gravity are not the same as those in general relativity. The modified gravity systems introduce additional terms that can significantly alter the stability conditions and can lead to deviations from the standard Zeldovich stability criterion \cite{, jasim2021anisotropic}. In order to completely comprehend stability in these scenarios, it is necessary to go further into the particular theory, even though the Zeldovich condition offers a starting point.} In terms of Zeldovich's criterion, any fluid sphere with a positive pressure-to-density ratio smaller than unity is regarded as physically acceptable \cite{l1962equation, zeldovich1971relativistic}. Accordingly, $\omega$ must always be continuous and fall between 0 and 1 throughout the stellar medium and this is satisfied by our solutions. \textcolor{blue}{Moreover, $\omega = 1/3$ represents an equation of state (EoS) for radiation for the matter constituent being relativistic \cite{das2016compact}. But in this study, we get several EoS parameter $\omega$ values other than $\omega = 1/3$. This suggests that our model is more appropriate for compact stars that radiate.}
 We have used a parametric plot to show the changes in pressure with respect to matter density in FIG.~\ref{eoss}. For various values of $\chi$, the variation in pressure-to-density ratio ($\omega$) is also shown in FIG.~\ref{eoss}.
 
\begin{table}[H]
\begin{center}
\caption{\label{table3} Numerically computed values central relativistic adiabatic index ($\Gamma_c$) and central EoS parameter ($\omega_c = p_c/\rho_c$) for the compact star Vela X-1 for different values of coupling constant $\chi$ (Taking $B=0.001 ~\rm{km}^{-2}$).}
\begin{tabular}{|c|c|c|c|c|}
\hline
$\chi$&  $\Gamma_c$ &$\omega_c = p_c/\rho_c$\\
\hline\hline
0.00& 4.06030 & 0.119786\\
 0.25& 4.04261 & 0.121119\\
 0.50& 4.02599 & 0.122416\\
 0.75& 4.01035 & 0.123677\\
 1.00& 3.99560 & 0.124903\\
 1.25& 3.98164 & 0.126094\\
\hline
\end{tabular}
\end{center}
\end{table}

\subsection{Mass-radius relationship, compactness and surface redshift}

\begin{figure}[H]
    \centering
        \includegraphics[scale=.45]{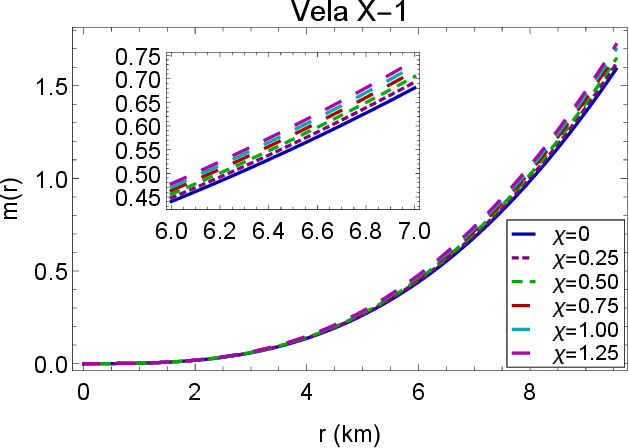}
        \includegraphics[scale=.45]{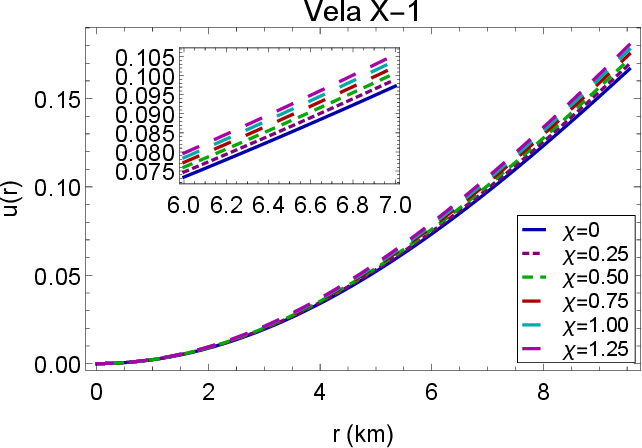}
        \includegraphics[scale=.45]{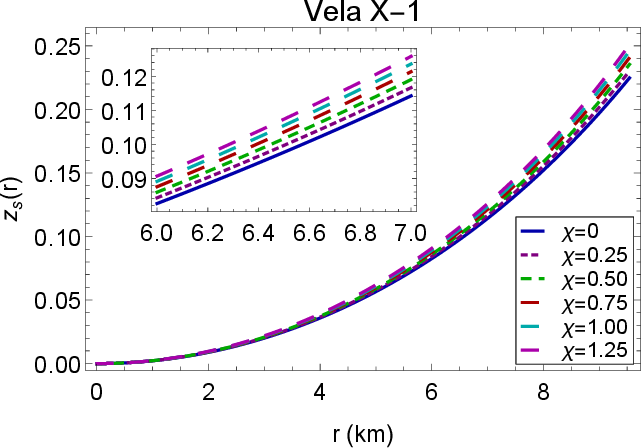}
       \caption{(left) The mass function, (middle) the compactness and (right) the surface redshift are plotted against the radius inside the stellar interior.\label{m11}}
\end{figure}

The mass function of the present system can be found as :
 \begin{eqnarray}
 m(r)&=&4\pi\int_0^r \xi^2 \rho^{\text{E}} (\xi)  d\xi=
\frac{3 B r^3 \big\{1 + 4 B r^2 + C \sqrt{1 + 4 B r^2}\big\}}{4 (1 + 5 B r^2 + 4 B^2 r^4)}.
 \end{eqnarray}
 \textcolor{blue}{The mass function at the core is regular, as seen in FIG.~\ref{m11}. It is simple to verify from FIG.~\ref{m11} that the mass function of a compact star is simply proportional to its radius, but it depends on the parameter $\chi$.} Also, we can see that the maximum mass is attained at the boundary of the star.
The compactness factor ($u$) is expressed as:
\begin{eqnarray}
u(r)=\frac{m(r)}{r}
\end{eqnarray} 
assigns the compact objects into the following groups: normal star ($u \sim 10^{-5}$), 
white dwarfs ($u~\sim 10^{-3}$), 
neutron star ($10^{-1} <~u < 1/4$), 
ultra-compact star ($1/4<~u<1/2$), and black hole ($u \sim 1/2$). 
Moreover, the surface redshift ($z_s$) can be computed using the formula below:

\begin{eqnarray}
z_s(r)&=&\frac{1}{\sqrt{1-2u}}-1,
\end{eqnarray}

\begin{table}[H]
\begin{center}
\caption{\label{table4} Numerically estimated values of mass, compactness factor, and surface redshift at the surface $r=\mathfrak{R}$ correspond to the compact star Vela X-1 for different values of coupling constant $\chi$ (Taking $B=0.001 ~\rm{km}^{-2}$).}
\begin{tabular}{|c|c|c|c|c|c|}
\hline
$\chi$&  $m(\mathfrak{R})$ & $u(\mathfrak{R})$ &$z_s(\mathfrak{R})$\\
\hline\hline
 0.00& 1.59851 & 0.167208 & 0.225741\\
 0.25& 1.62636 & 0.170122 & 0.231142\\
 0.50& 1.65351 & 0.172962 & 0.236476\\
 0.75& 1.67999 & 0.175731 & 0.241746\\
 1.00& 1.70582 & 0.178434 & 0.246952\\
 1.25& 1.73103 & 0.18107  & 0.252096\\
\hline
\end{tabular}
\end{center}
\end{table}

FIG.~\ref{m11} displays the profiles of compactness and surface redshift. It is shown that they are all monotonically increasing with respect to $r$ confirming the viability of our presented model.

\begin{figure}[H]
    \centering
        \includegraphics[scale=.45]{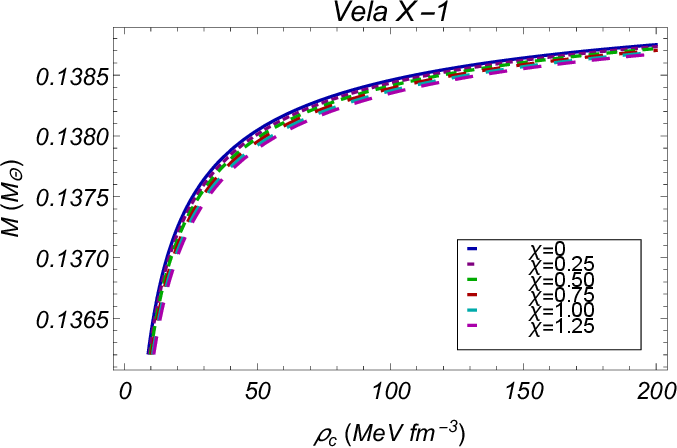}
        \includegraphics[scale=.45]{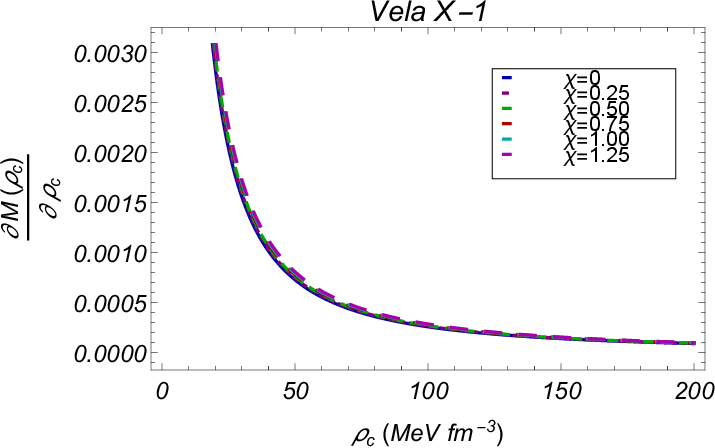}
        \caption{(left) Total mass versus central density and (right) $\frac{\partial M}{\partial \rho_c}$ versus central density plots for the chosen compact stellar configuration.\label{stab}}
\end{figure}

\textcolor{purple}{\subsection{Stability Test through Harrison-Zeldovich-Novikov criterion} \label{stable}
We are particularly interested in studying the stability of the compact stellar configuration. The Harrison-Zeldovich-Novikov criterion \cite{harrison1965gravitation, zeldovich1971relativistic} describes the stability of a compact object. The necessary (but not sufficient) condition for the stability of a compact stellar configuration is that the total gravitational mass ($M$) of a compact star should be an increasing function of the central density ($\rho_c$) throughout the stellar region.  Mathematically defined as, $\frac{\partial M}{\partial \rho_c} > 0$ \cite{Glendenning:1998ag, Arbanil:2016wud}. This condition, well beyond the purview of GR, has been commonly used in the literature of modified gravity theories also because of its simplicity. The dependency of compact star's total mass in gravitational units on their central density is depicted in FIG.~\ref{stab} and we can verify that the total mass ($M (\rho_c)$) increases with increasing central density ($\rho_c$). Also the right panel of FIG.~\ref{stab} describes the positive behaviour of $\frac{\partial M}{\partial \rho_c}$ with respect to central density ($\rho_c$). Both curves exhibit asymptotic behaviour. Thus we can conclude that our current model satisfies
Harrison-Zeldovich-Novikov stability criterion and maintains stable equilibrium throughout the region.}

\section{Discussion}
Our main work has consisted of building an isotropic star in the $f(R,\, T)$ gravity with baryonic matter which is static, symmetric, and free of singularities, and then we have studied its nature. For this theoretical investigation, we have chosen a particular compact star, Vela X-1, and we have given our findings conceptually as well as pictorially in this study.
We make the following assumptions throughout the work: $p(r= \mathfrak{R})=0$ and the continuity of the metric functions $g_{tt}$ and $g_{rr}$ over the boundary surface $r= \mathfrak{R}$. The Heintzmann IIa metric coefficients contain constants that one must find by matching the exterior Schwarzschild metric with the interior metric. Furthermore, the physical attributes are plotted for values of the coupling parameter $\chi = 0.00, 0.25, 0.50, 0.75, 1.00$, and $1.25$. The graphical representations and tables that conform to point out the main remarkable conclusions which can be listed as follows:

\begin{itemize}

\item The Heintzmann IIa metric functions, which rely on arbitrary constants $A, B$, and $C$, have been selected as the basis for the development of this star model. 
The values of these above arbitrary constants can be correctly inferred from the boundary conditions, which are given in Table \ref{tb12}, using the estimated mass and radius of the star under study. Interestingly, with the increase of $\chi$, $A$ remains unchanged, that is, independent of the choice of $\chi$, but $C$ increases.

\item The graphic depictions of $e^{\nu(r)}$ and $e^{\lambda(r)}$ in FIG.\ref{metric} indicate that they are finite and non-singular over the radius of stars for variable $\chi$ which promises that within the context of $f(R, T)$ gravity, these metric potentials are appropriate for producing a non-singular model for celestial compact stars.

\item FIG. \ref{pp} shows that, across a range of $\chi$ values, the energy density $\rho(r)$ and pressure $p(r)$ inside the star remain continuous, positively finite, and exhibit a smooth decreasing tendency towards the surface for varying $\chi$. So, this compact star structure has well-defined pressure and density as physical characteristics. At the boundary of the stellar structure, the pressure disappears, while the density remains positive. Moreover, while $\chi $ increases from $0$ to $1.25$, the values of $\rho(r)$ and $p(r)$ decrease, meaning that compact stars have higher pressure and energy density in standard Einstein's gravity than in modified $f(R,\, T)$ gravity. Hence, compared to ordinary Einstein's gravity, modified $f(R,\, T)$ gravity is more suited for providing ultimate stable compact stars in Heintzmann IIa spacetime. In addition, we evaluated the values of $\rho_c$, $\rho_s$, and $p_c$, which rely on $\chi$, as shown in Table \ref{tb1}. We conclude that $\rho_c$, $\rho_s$, and $p_c$ are all positive in the stellar medium based on this table.

\item As the value of $r$ goes from center to border, the gradient components of matter-energy density and pressure in FIG. \ref{grad5} show a negative and monotonically decreasing trend, moving from zero to a negative region for various values of $\chi$.

\item As illustrated visually in FIG. \ref{ener1}, our consequent model satisfies all four energy criteria, namely the NEC, WEC, SEC, and DEC, for various values of $\chi$. Moreover, these are constantly positive throughout the entire star, indicating that our proposed approach is viable.

\item Next, we examine the causality criterion, one of the essential stability mechanisms, to ensure the consistency of the model. From FIG. \ref{sv}(Left), we can see that the square of the sound velocity $V^2$ lies within $(0,1)$ within the stellar body, indicating that the model is physically stable.

\item We analyze the behavior of the adiabatic index, $\Gamma$, in FIG. \ref{sv} (Right). The graph illustrates that it takes values greater than $4/3$ throughout the fluid sphere, indicating that the stability requirements are met. Moreover, Table \ref{table3} displays the numerical values of the central relativistic adiabatic index $\Gamma_c$ which allows us to verify that $\Gamma > 4/3$.

\item In the context of $f(R,\, T)$ gravity, the isotropic matter configurations always remain in equilibrium in the system for every given value of $\chi$, as demonstrated by the forces related to the TOV equation in FIG. \ref{tov1}. In this scenario, the additional force resulting from modified gravity ($F_m$) and the hydrostatic force ($F_h$) are repulsive, while the gravitational force ($F_g$) is attractive. As a result, the matter configurations represented by the solutions are in an equilibrium condition that prevents gravitational collapse and causes the formation of stable compact objects.

\item 
For a compact stellar object to be stable, determining the equation of state, i.e. the relationship between pressure and density, is very significant. The change in pressure with respect to density and the pressure-density ratio ($\omega$) for various values of $\chi$ are shown in FIG. \ref{eoss}. This illustration shows that our fluid model satisfies the Zeldovich condition, that is, the pressure-to-density ratio falls between 0 and 1 \cite{zeldovich1971relativistic}. In Table \ref{table3}, we have tabulated the values of the central EoS parameter ($\omega_c$) to verify this fact.

\item FIG. \ref{m11} displays the variations of the mass function, compactness factor, and surface redshift for different values of $\chi$ within compact object formations. From figure, we see that they are all well-behaved, increasing steadily with $r$ and attain their peak values at the boundary. We have estimated the values of mass function, compactness factor, and surface redshift at the stellar surface in Table \ref{table4}.
\textcolor{purple}{ \item The left panel of FIG.~\ref{stab} shows how the total mass varies with central density. While the right panel exhibits the feature $\frac{\partial M}{\partial \rho_c} > 0$ for our system and demonstrates system stability by adhering to the Harrison-Zeldovich-Novikov stability criterion. }
 
In summary, our suggested strange star model, derived from the isotropic Heintzmann IIa metric solution in $f(R,\, T)$ gravity theory, is singularity-free and meets all the requirements of a stable and acceptable model scientifically. In the astrophysical scenario, it seems that our representation could be successful in playing a greater extent.

\end{itemize}
%%%%%%%%%%%%%%%%%%%%%%%%%%%

\section*{Author contributions}
\textbf{Pramit Rej} contributed to conceptualization, validation, investigation, computer code design for data analysis, original draft preparation, overall supervision, writing - review \& editing of the project.
\textbf{Akashdip Karmakar} performed validation, mathematical analysis, writing - review, \& editing. Both authors read and approved the entire manuscript.

\section*{Acknowledgements} 
\textbf{Pramit Rej} is thankful to the Inter-University Centre for Astronomy and Astrophysics (IUCAA), Pune, Government of India, for providing Visiting Associateship.

\section*{Declarations}
\textbf{Funding:} The authors did not receive any funding in the form of financial aid or grant from any institution or organization for the present research work.\par
\textbf{Data Availability Statement:} The results are obtained
through purely theoretical calculations and can be verified analytically;
thus this manuscript has no associated data, or the data will not be deposited. \par
\textbf{Conflicts of Interest:} The authors declare that they have no known competing financial interests or personal relationships that could have appeared to influence the work reported in this paper.\par

\bibliographystyle{apsrev4-1}
\bibliography{ref1}

\begin{thebibliography}{76}%
\makeatletter
\providecommand \@ifxundefined [1]{%
 \@ifx{#1\undefined}
}%
\providecommand \@ifnum [1]{%
 \ifnum #1\expandafter \@firstoftwo
 \else \expandafter \@secondoftwo
 \fi
}%
\providecommand \@ifx [1]{%
 \ifx #1\expandafter \@firstoftwo
 \else \expandafter \@secondoftwo
 \fi
}%
\providecommand \natexlab [1]{#1}%
\providecommand \enquote  [1]{``#1''}%
\providecommand \bibnamefont  [1]{#1}%
\providecommand \bibfnamefont [1]{#1}%
\providecommand \citenamefont [1]{#1}%
\providecommand \href@noop [0]{\@secondoftwo}%
\providecommand \href [0]{\begingroup \@sanitize@url \@href}%
\providecommand \@href[1]{\@@startlink{#1}\@@href}%
\providecommand \@@href[1]{\endgroup#1\@@endlink}%
\providecommand \@sanitize@url [0]{\catcode `\\12\catcode `\$12\catcode `\&12\catcode `\#12\catcode `\^12\catcode `\_12\catcode `\%12\relax}%
\providecommand \@@startlink[1]{}%
\providecommand \@@endlink[0]{}%
\providecommand \url  [0]{\begingroup\@sanitize@url \@url }%
\providecommand \@url [1]{\endgroup\@href {#1}{\urlprefix }}%
\providecommand \urlprefix  [0]{URL }%
\providecommand \Eprint [0]{\href }%
\providecommand \doibase [0]{http://dx.doi.org/}%
\providecommand \selectlanguage [0]{\@gobble}%
\providecommand \bibinfo  [0]{\@secondoftwo}%
\providecommand \bibfield  [0]{\@secondoftwo}%
\providecommand \translation [1]{[#1]}%
\providecommand \BibitemOpen [0]{}%
\providecommand \bibitemStop [0]{}%
\providecommand \bibitemNoStop [0]{.\EOS\space}%
\providecommand \EOS [0]{\spacefactor3000\relax}%
\providecommand \BibitemShut  [1]{\csname bibitem#1\endcsname}%
\let\auto@bib@innerbib\@empty
%</preamble>
\bibitem [{\citenamefont {Schwarzschild}(1916)}]{schwarzschild1916gravitationsfeld1}%
  \BibitemOpen
  \bibfield  {author} {\bibinfo {author} {\bibfnamefont {K.}~\bibnamefont {Schwarzschild}},\ }\href@noop {} {\bibfield  {journal} {\bibinfo  {journal} {Sitzungsber. Preuss. Akad. Wiss. Berlin (Math. Phys. )}\ }\textbf {\bibinfo {volume} {1916}},\ \bibinfo {pages} {189} (\bibinfo {year} {1916})},\ \Eprint {http://arxiv.org/abs/physics/9905030} {arXiv:physics/9905030} \BibitemShut {NoStop}%
\bibitem [{\citenamefont {Tolman}(1939)}]{tolman1939static}%
  \BibitemOpen
  \bibfield  {author} {\bibinfo {author} {\bibfnamefont {R.~C.}\ \bibnamefont {Tolman}},\ }\href {https://doi.org/10.1103/PhysRev.55.364} {\bibfield  {journal} {\bibinfo  {journal} {Physical Review}\ }\textbf {\bibinfo {volume} {55}},\ \bibinfo {pages} {364} (\bibinfo {year} {1939})}\BibitemShut {NoStop}%
\bibitem [{\citenamefont {Riess}\ \emph {et~al.}(1998)\citenamefont {Riess}, \citenamefont {Filippenko}, \citenamefont {Challis}, \citenamefont {Clocchiatti}, \citenamefont {Diercks}, \citenamefont {Garnavich}, \citenamefont {Gilliland}, \citenamefont {Hogan}, \citenamefont {Jha}, \citenamefont {Kirshner} \emph {et~al.}}]{riess1998observational}%
  \BibitemOpen
  \bibfield  {author} {\bibinfo {author} {\bibfnamefont {A.~G.}\ \bibnamefont {Riess}}, \bibinfo {author} {\bibfnamefont {A.~V.}\ \bibnamefont {Filippenko}}, \bibinfo {author} {\bibfnamefont {P.}~\bibnamefont {Challis}}, \bibinfo {author} {\bibfnamefont {A.}~\bibnamefont {Clocchiatti}}, \bibinfo {author} {\bibfnamefont {A.}~\bibnamefont {Diercks}}, \bibinfo {author} {\bibfnamefont {P.~M.}\ \bibnamefont {Garnavich}}, \bibinfo {author} {\bibfnamefont {R.~L.}\ \bibnamefont {Gilliland}}, \bibinfo {author} {\bibfnamefont {C.~J.}\ \bibnamefont {Hogan}}, \bibinfo {author} {\bibfnamefont {S.}~\bibnamefont {Jha}}, \bibinfo {author} {\bibfnamefont {R.~P.}\ \bibnamefont {Kirshner}},  \emph {et~al.},\ }\href {\doibase 10.1086/300499} {\bibfield  {journal} {\bibinfo  {journal} {The astronomical journal}\ }\textbf {\bibinfo {volume} {116}},\ \bibinfo {pages} {1009} (\bibinfo {year} {1998})}\BibitemShut {NoStop}%
\bibitem [{\citenamefont {Koyama}(2016)}]{koyama2016cosmological}%
  \BibitemOpen
  \bibfield  {author} {\bibinfo {author} {\bibfnamefont {K.}~\bibnamefont {Koyama}},\ }\href {\doibase 10.1088/0034-4885/79/4/046902} {\bibfield  {journal} {\bibinfo  {journal} {Reports on Progress in Physics}\ }\textbf {\bibinfo {volume} {79}},\ \bibinfo {pages} {046902} (\bibinfo {year} {2016})}\BibitemShut {NoStop}%
\bibitem [{\citenamefont {Copeland}\ \emph {et~al.}(2006)\citenamefont {Copeland}, \citenamefont {Sami},\ and\ \citenamefont {Tsujikawa}}]{copeland2006dynamics}%
  \BibitemOpen
  \bibfield  {author} {\bibinfo {author} {\bibfnamefont {E.~J.}\ \bibnamefont {Copeland}}, \bibinfo {author} {\bibfnamefont {M.}~\bibnamefont {Sami}}, \ and\ \bibinfo {author} {\bibfnamefont {S.}~\bibnamefont {Tsujikawa}},\ }\href {\doibase 10.1142/S021827180600942X} {\bibfield  {journal} {\bibinfo  {journal} {International Journal of Modern Physics D}\ }\textbf {\bibinfo {volume} {15}},\ \bibinfo {pages} {1753} (\bibinfo {year} {2006})}\BibitemShut {NoStop}%
\bibitem [{\citenamefont {Joyce}\ \emph {et~al.}(2015)\citenamefont {Joyce}, \citenamefont {Jain}, \citenamefont {Khoury},\ and\ \citenamefont {Trodden}}]{joyce2015beyond}%
  \BibitemOpen
  \bibfield  {author} {\bibinfo {author} {\bibfnamefont {A.}~\bibnamefont {Joyce}}, \bibinfo {author} {\bibfnamefont {B.}~\bibnamefont {Jain}}, \bibinfo {author} {\bibfnamefont {J.}~\bibnamefont {Khoury}}, \ and\ \bibinfo {author} {\bibfnamefont {M.}~\bibnamefont {Trodden}},\ }\href {\doibase https://doi.org/10.1016/j.physrep.2014.12.002} {\enquote {\bibinfo {title} {Beyond the cosmological standard model. phys rept 568: 1--98},}\ } (\bibinfo {year} {2015})\BibitemShut {NoStop}%
\bibitem [{\citenamefont {Verde}\ \emph {et~al.}(2003)\citenamefont {Verde} \emph {et~al.}}]{WMAP:2003pyh}%
  \BibitemOpen
  \bibfield  {author} {\bibinfo {author} {\bibfnamefont {L.}~\bibnamefont {Verde}} \emph {et~al.} (\bibinfo {collaboration} {WMAP}),\ }\href {\doibase 10.1086/377335} {\bibfield  {journal} {\bibinfo  {journal} {Astrophys. J. Suppl.}\ }\textbf {\bibinfo {volume} {148}},\ \bibinfo {pages} {195} (\bibinfo {year} {2003})},\ \Eprint {http://arxiv.org/abs/astro-ph/0302218} {arXiv:astro-ph/0302218} \BibitemShut {NoStop}%
\bibitem [{\citenamefont {Komatsu}\ \emph {et~al.}(2011)\citenamefont {Komatsu} \emph {et~al.}}]{WMAP:2010qai}%
  \BibitemOpen
  \bibfield  {author} {\bibinfo {author} {\bibfnamefont {E.}~\bibnamefont {Komatsu}} \emph {et~al.} (\bibinfo {collaboration} {WMAP}),\ }\href {\doibase 10.1088/0067-0049/192/2/18} {\bibfield  {journal} {\bibinfo  {journal} {Astrophys. J. Suppl.}\ }\textbf {\bibinfo {volume} {192}},\ \bibinfo {pages} {18} (\bibinfo {year} {2011})},\ \Eprint {http://arxiv.org/abs/1001.4538} {arXiv:1001.4538 [astro-ph.CO]} \BibitemShut {NoStop}%
\bibitem [{\citenamefont {Nojiri}\ \emph {et~al.}(2016)\citenamefont {Nojiri}, \citenamefont {Odintsov},\ and\ \citenamefont {Oikonomou}}]{nojiri2016unimodular}%
  \BibitemOpen
  \bibfield  {author} {\bibinfo {author} {\bibfnamefont {S.}~\bibnamefont {Nojiri}}, \bibinfo {author} {\bibfnamefont {S.~D.}\ \bibnamefont {Odintsov}}, \ and\ \bibinfo {author} {\bibfnamefont {V.}~\bibnamefont {Oikonomou}},\ }\href {\doibase 10.1088/1475-7516/2016/05/046} {\bibfield  {journal} {\bibinfo  {journal} {Journal of Cosmology and Astroparticle Physics}\ }\textbf {\bibinfo {volume} {2016}},\ \bibinfo {pages} {046} (\bibinfo {year} {2016})}\BibitemShut {NoStop}%
\bibitem [{\citenamefont {Garc{\'\i}a-Aspeitia}\ \emph {et~al.}(2019)\citenamefont {Garc{\'\i}a-Aspeitia}, \citenamefont {Mart{\'\i}nez-Robles}, \citenamefont {Hern{\'a}ndez-Almada}, \citenamefont {Maga{\~n}a},\ and\ \citenamefont {Motta}}]{garcia2019cosmic}%
  \BibitemOpen
  \bibfield  {author} {\bibinfo {author} {\bibfnamefont {M.~A.}\ \bibnamefont {Garc{\'\i}a-Aspeitia}}, \bibinfo {author} {\bibfnamefont {C.}~\bibnamefont {Mart{\'\i}nez-Robles}}, \bibinfo {author} {\bibfnamefont {A.}~\bibnamefont {Hern{\'a}ndez-Almada}}, \bibinfo {author} {\bibfnamefont {J.}~\bibnamefont {Maga{\~n}a}}, \ and\ \bibinfo {author} {\bibfnamefont {V.}~\bibnamefont {Motta}},\ }\href {\doibase https://doi.org/10.1103/PhysRevD.99.123525} {\bibfield  {journal} {\bibinfo  {journal} {Physical Review D}\ }\textbf {\bibinfo {volume} {99}},\ \bibinfo {pages} {123525} (\bibinfo {year} {2019})}\BibitemShut {NoStop}%
\bibitem [{\citenamefont {Nojiri}\ and\ \citenamefont {Odintsov}(2007)}]{nojiri2007introduction}%
  \BibitemOpen
  \bibfield  {author} {\bibinfo {author} {\bibfnamefont {S.}~\bibnamefont {Nojiri}}\ and\ \bibinfo {author} {\bibfnamefont {S.~D.}\ \bibnamefont {Odintsov}},\ }\href {\doibase https://doi.org/10.1142/S0219887807001928} {\bibfield  {journal} {\bibinfo  {journal} {International Journal of Geometric Methods in Modern Physics}\ }\textbf {\bibinfo {volume} {4}},\ \bibinfo {pages} {115} (\bibinfo {year} {2007})}\BibitemShut {NoStop}%
\bibitem [{\citenamefont {Nojiri}\ and\ \citenamefont {Odintsov}(2011)}]{nojiri2011unified}%
  \BibitemOpen
  \bibfield  {author} {\bibinfo {author} {\bibfnamefont {S.}~\bibnamefont {Nojiri}}\ and\ \bibinfo {author} {\bibfnamefont {S.~D.}\ \bibnamefont {Odintsov}},\ }\href {\doibase https://doi.org/10.1016/j.physrep.2011.04.001} {\bibfield  {journal} {\bibinfo  {journal} {Physics Reports}\ }\textbf {\bibinfo {volume} {505}},\ \bibinfo {pages} {59} (\bibinfo {year} {2011})}\BibitemShut {NoStop}%
\bibitem [{\citenamefont {Xu}\ \emph {et~al.}(2019)\citenamefont {Xu}, \citenamefont {Li}, \citenamefont {Harko},\ and\ \citenamefont {Liang}}]{xu2019regular}%
  \BibitemOpen
  \bibfield  {author} {\bibinfo {author} {\bibfnamefont {Y.}~\bibnamefont {Xu}}, \bibinfo {author} {\bibfnamefont {G.}~\bibnamefont {Li}}, \bibinfo {author} {\bibfnamefont {T.}~\bibnamefont {Harko}}, \ and\ \bibinfo {author} {\bibfnamefont {S.-D.}\ \bibnamefont {Liang}},\ }\href {\doibase https://doi.org/10.1140/epjc/s10052-019-7207-4} {\bibfield  {journal} {\bibinfo  {journal} {Eur. Phys. J. C}\ }\textbf {\bibinfo {volume} {79}},\ \bibinfo {pages} {708} (\bibinfo {year} {2019})}\BibitemShut {NoStop}%
\bibitem [{\citenamefont {Arora}\ \emph {et~al.}(2020)\citenamefont {Arora}, \citenamefont {Pacif}, \citenamefont {Bhattacharjee},\ and\ \citenamefont {Sahoo}}]{arora2020f}%
  \BibitemOpen
  \bibfield  {author} {\bibinfo {author} {\bibfnamefont {S.}~\bibnamefont {Arora}}, \bibinfo {author} {\bibfnamefont {S.}~\bibnamefont {Pacif}}, \bibinfo {author} {\bibfnamefont {S.}~\bibnamefont {Bhattacharjee}}, \ and\ \bibinfo {author} {\bibfnamefont {P.}~\bibnamefont {Sahoo}},\ }\href {\doibase https://doi.org/10.1016/j.dark.2020.100664} {\bibfield  {journal} {\bibinfo  {journal} {Physics of the Dark Universe}\ }\textbf {\bibinfo {volume} {30}},\ \bibinfo {pages} {100664} (\bibinfo {year} {2020})}\BibitemShut {NoStop}%
\bibitem [{\citenamefont {Bahamonde}\ \emph {et~al.}(2023)\citenamefont {Bahamonde}, \citenamefont {Dialektopoulos}, \citenamefont {Escamilla-Rivera}, \citenamefont {Farrugia}, \citenamefont {Gakis}, \citenamefont {Hendry}, \citenamefont {Hohmann}, \citenamefont {Said}, \citenamefont {Mifsud},\ and\ \citenamefont {Di~Valentino}}]{bahamonde2023teleparallel}%
  \BibitemOpen
  \bibfield  {author} {\bibinfo {author} {\bibfnamefont {S.}~\bibnamefont {Bahamonde}}, \bibinfo {author} {\bibfnamefont {K.~F.}\ \bibnamefont {Dialektopoulos}}, \bibinfo {author} {\bibfnamefont {C.}~\bibnamefont {Escamilla-Rivera}}, \bibinfo {author} {\bibfnamefont {G.}~\bibnamefont {Farrugia}}, \bibinfo {author} {\bibfnamefont {V.}~\bibnamefont {Gakis}}, \bibinfo {author} {\bibfnamefont {M.}~\bibnamefont {Hendry}}, \bibinfo {author} {\bibfnamefont {M.}~\bibnamefont {Hohmann}}, \bibinfo {author} {\bibfnamefont {J.~L.}\ \bibnamefont {Said}}, \bibinfo {author} {\bibfnamefont {J.}~\bibnamefont {Mifsud}}, \ and\ \bibinfo {author} {\bibfnamefont {E.}~\bibnamefont {Di~Valentino}},\ }\href {\doibase 10.1088/1361-6633/ac9cef} {\bibfield  {journal} {\bibinfo  {journal} {Reports on Progress in Physics}\ }\textbf {\bibinfo {volume} {86}},\ \bibinfo {pages} {026901} (\bibinfo {year} {2023})}\BibitemShut {NoStop}%
\bibitem [{\citenamefont {Atazadeh}\ and\ \citenamefont {Darabi}(2014)}]{atazadeh2014energy}%
  \BibitemOpen
  \bibfield  {author} {\bibinfo {author} {\bibfnamefont {K.}~\bibnamefont {Atazadeh}}\ and\ \bibinfo {author} {\bibfnamefont {F.}~\bibnamefont {Darabi}},\ }\href {\doibase https://doi.org/10.1007/s10714-014-1664-8} {\bibfield  {journal} {\bibinfo  {journal} {General Relativity and Gravitation}\ }\textbf {\bibinfo {volume} {46}},\ \bibinfo {pages} {1} (\bibinfo {year} {2014})}\BibitemShut {NoStop}%
\bibitem [{\citenamefont {De~Felice}\ \emph {et~al.}(2011)\citenamefont {De~Felice}, \citenamefont {Suyama},\ and\ \citenamefont {Tanaka}}]{de2011stability}%
  \BibitemOpen
  \bibfield  {author} {\bibinfo {author} {\bibfnamefont {A.}~\bibnamefont {De~Felice}}, \bibinfo {author} {\bibfnamefont {T.}~\bibnamefont {Suyama}}, \ and\ \bibinfo {author} {\bibfnamefont {T.}~\bibnamefont {Tanaka}},\ }\href {\doibase https://doi.org/10.1103/PhysRevD.83.104035} {\bibfield  {journal} {\bibinfo  {journal} {Physical Review D}\ }\textbf {\bibinfo {volume} {83}},\ \bibinfo {pages} {104035} (\bibinfo {year} {2011})}\BibitemShut {NoStop}%
\bibitem [{\citenamefont {Harko}\ \emph {et~al.}(2011)\citenamefont {Harko}, \citenamefont {Lobo}, \citenamefont {Nojiri},\ and\ \citenamefont {Odintsov}}]{harko2011f}%
  \BibitemOpen
  \bibfield  {author} {\bibinfo {author} {\bibfnamefont {T.}~\bibnamefont {Harko}}, \bibinfo {author} {\bibfnamefont {F.~S.~N.}\ \bibnamefont {Lobo}}, \bibinfo {author} {\bibfnamefont {S.}~\bibnamefont {Nojiri}}, \ and\ \bibinfo {author} {\bibfnamefont {S.~D.}\ \bibnamefont {Odintsov}},\ }\href {\doibase 10.1103/PhysRevD.84.024020} {\bibfield  {journal} {\bibinfo  {journal} {Physical Review D}\ }\textbf {\bibinfo {volume} {84}},\ \bibinfo {pages} {024020} (\bibinfo {year} {2011})}\BibitemShut {NoStop}%
\bibitem [{\citenamefont {Chakraborty}(2013)}]{chakraborty2013alternative}%
  \BibitemOpen
  \bibfield  {author} {\bibinfo {author} {\bibfnamefont {S.}~\bibnamefont {Chakraborty}},\ }\href {\doibase https://doi.org/10.1007/s10714-013-1577-y} {\bibfield  {journal} {\bibinfo  {journal} {General Relativity and Gravitation}\ }\textbf {\bibinfo {volume} {45}},\ \bibinfo {pages} {2039} (\bibinfo {year} {2013})}\BibitemShut {NoStop}%
\bibitem [{\citenamefont {Houndjo}(2012)}]{doi:10.1142/S0218271812500034}%
  \BibitemOpen
  \bibfield  {author} {\bibinfo {author} {\bibfnamefont {M.~J.~S.}\ \bibnamefont {Houndjo}},\ }\href {\doibase 10.1142/S0218271812500034} {\bibfield  {journal} {\bibinfo  {journal} {International Journal of Modern Physics D}\ }\textbf {\bibinfo {volume} {21}},\ \bibinfo {pages} {1250003} (\bibinfo {year} {2012})}\BibitemShut {NoStop}%
\bibitem [{\citenamefont {Sharif}\ and\ \citenamefont {Zubair}(2013)}]{sharif2013energy}%
  \BibitemOpen
  \bibfield  {author} {\bibinfo {author} {\bibfnamefont {M.}~\bibnamefont {Sharif}}\ and\ \bibinfo {author} {\bibfnamefont {M.}~\bibnamefont {Zubair}},\ }\href {\doibase https://doi.org/10.1007/JHEP12(2013)079} {\bibfield  {journal} {\bibinfo  {journal} {Journal of High Energy Physics}\ }\textbf {\bibinfo {volume} {2013}},\ \bibinfo {pages} {1} (\bibinfo {year} {2013})}\BibitemShut {NoStop}%
\bibitem [{\citenamefont {Odintsov}\ and\ \citenamefont {S{\'a}ez-G{\'o}mez}(2013)}]{odintsov2013f}%
  \BibitemOpen
  \bibfield  {author} {\bibinfo {author} {\bibfnamefont {S.~D.}\ \bibnamefont {Odintsov}}\ and\ \bibinfo {author} {\bibfnamefont {D.}~\bibnamefont {S{\'a}ez-G{\'o}mez}},\ }\href {\doibase https://doi.org/10.1016/j.physletb.2013.07.026} {\bibfield  {journal} {\bibinfo  {journal} {Physics Letters B}\ }\textbf {\bibinfo {volume} {725}},\ \bibinfo {pages} {437} (\bibinfo {year} {2013})}\BibitemShut {NoStop}%
\bibitem [{\citenamefont {Shabani}\ and\ \citenamefont {Farhoudi}(2014{\natexlab{a}})}]{PhysRevD.90.044031}%
  \BibitemOpen
  \bibfield  {author} {\bibinfo {author} {\bibfnamefont {H.}~\bibnamefont {Shabani}}\ and\ \bibinfo {author} {\bibfnamefont {M.}~\bibnamefont {Farhoudi}},\ }\href {\doibase https://doi.org/10.1103/PhysRevD.90.044031} {\bibfield  {journal} {\bibinfo  {journal} {Phys. Rev. D}\ }\textbf {\bibinfo {volume} {90}},\ \bibinfo {pages} {044031} (\bibinfo {year} {2014}{\natexlab{a}})}\BibitemShut {NoStop}%
\bibitem [{\citenamefont {Shabani}\ and\ \citenamefont {Farhoudi}(2013)}]{Shabani:2013djy}%
  \BibitemOpen
  \bibfield  {author} {\bibinfo {author} {\bibfnamefont {H.}~\bibnamefont {Shabani}}\ and\ \bibinfo {author} {\bibfnamefont {M.}~\bibnamefont {Farhoudi}},\ }\href {\doibase 10.1103/PhysRevD.88.044048} {\bibfield  {journal} {\bibinfo  {journal} {Phys. Rev. D}\ }\textbf {\bibinfo {volume} {88}},\ \bibinfo {pages} {044048} (\bibinfo {year} {2013})},\ \Eprint {http://arxiv.org/abs/1306.3164} {arXiv:1306.3164 [gr-qc]} \BibitemShut {NoStop}%
\bibitem [{\citenamefont {Shabani}\ and\ \citenamefont {Farhoudi}(2014{\natexlab{b}})}]{Shabani:2014xvi}%
  \BibitemOpen
  \bibfield  {author} {\bibinfo {author} {\bibfnamefont {H.}~\bibnamefont {Shabani}}\ and\ \bibinfo {author} {\bibfnamefont {M.}~\bibnamefont {Farhoudi}},\ }\href {\doibase 10.1103/PhysRevD.90.044031} {\bibfield  {journal} {\bibinfo  {journal} {Phys. Rev. D}\ }\textbf {\bibinfo {volume} {90}},\ \bibinfo {pages} {044031} (\bibinfo {year} {2014}{\natexlab{b}})},\ \Eprint {http://arxiv.org/abs/1407.6187} {arXiv:1407.6187 [gr-qc]} \BibitemShut {NoStop}%
\bibitem [{\citenamefont {dos Santos~Jr}\ \emph {et~al.}(2019)\citenamefont {dos Santos~Jr}, \citenamefont {Carvalho}, \citenamefont {Moraes}, \citenamefont {Lenzi},\ and\ \citenamefont {Malheiro}}]{dos2019conservative}%
  \BibitemOpen
  \bibfield  {author} {\bibinfo {author} {\bibfnamefont {S.~I.}\ \bibnamefont {dos Santos~Jr}}, \bibinfo {author} {\bibfnamefont {G.~A.}\ \bibnamefont {Carvalho}}, \bibinfo {author} {\bibfnamefont {P.~H. R.~S.}\ \bibnamefont {Moraes}}, \bibinfo {author} {\bibfnamefont {C.~H.}\ \bibnamefont {Lenzi}}, \ and\ \bibinfo {author} {\bibfnamefont {M.}~\bibnamefont {Malheiro}},\ }\href {\doibase https://doi.org/10.1140/epjp/i2019-12830-8} {\bibfield  {journal} {\bibinfo  {journal} {The European Physical Journal Plus}\ }\textbf {\bibinfo {volume} {134}},\ \bibinfo {pages} {398} (\bibinfo {year} {2019})}\BibitemShut {NoStop}%
\bibitem [{\citenamefont {Moraes}\ \emph {et~al.}(2016)\citenamefont {Moraes}, \citenamefont {Arba{\~n}il},\ and\ \citenamefont {Malheiro}}]{moraes2016stellar}%
  \BibitemOpen
  \bibfield  {author} {\bibinfo {author} {\bibfnamefont {P.}~\bibnamefont {Moraes}}, \bibinfo {author} {\bibfnamefont {J.~D.}\ \bibnamefont {Arba{\~n}il}}, \ and\ \bibinfo {author} {\bibfnamefont {M.}~\bibnamefont {Malheiro}},\ }\href {\doibase 10.1088/1475-7516/2016/06/005} {\bibfield  {journal} {\bibinfo  {journal} {Journal of Cosmology and Astroparticle Physics}\ }\textbf {\bibinfo {volume} {2016}},\ \bibinfo {pages} {005} (\bibinfo {year} {2016})}\BibitemShut {NoStop}%
\bibitem [{\citenamefont {Sun}\ and\ \citenamefont {Huang}(2016)}]{sun2016cosmology}%
  \BibitemOpen
  \bibfield  {author} {\bibinfo {author} {\bibfnamefont {G.}~\bibnamefont {Sun}}\ and\ \bibinfo {author} {\bibfnamefont {Y.-C.}\ \bibnamefont {Huang}},\ }\href {\doibase https://doi.org/10.1142/S0218271816500383} {\bibfield  {journal} {\bibinfo  {journal} {International Journal of Modern Physics D}\ }\textbf {\bibinfo {volume} {25}},\ \bibinfo {pages} {1650038} (\bibinfo {year} {2016})}\BibitemShut {NoStop}%
\bibitem [{\citenamefont {Zaregonbadi}\ \emph {et~al.}(2016)\citenamefont {Zaregonbadi}, \citenamefont {Farhoudi},\ and\ \citenamefont {Riazi}}]{zaregonbadi2016dark}%
  \BibitemOpen
  \bibfield  {author} {\bibinfo {author} {\bibfnamefont {R.}~\bibnamefont {Zaregonbadi}}, \bibinfo {author} {\bibfnamefont {M.}~\bibnamefont {Farhoudi}}, \ and\ \bibinfo {author} {\bibfnamefont {N.}~\bibnamefont {Riazi}},\ }\href {\doibase https://doi.org/10.1103/PhysRevD.94.084052} {\bibfield  {journal} {\bibinfo  {journal} {Physical Review D}\ }\textbf {\bibinfo {volume} {94}},\ \bibinfo {pages} {084052} (\bibinfo {year} {2016})}\BibitemShut {NoStop}%
\bibitem [{\citenamefont {Moraes}\ and\ \citenamefont {Sahoo}(2019)}]{moraes2019wormholes}%
  \BibitemOpen
  \bibfield  {author} {\bibinfo {author} {\bibfnamefont {P.}~\bibnamefont {Moraes}}\ and\ \bibinfo {author} {\bibfnamefont {P.}~\bibnamefont {Sahoo}},\ }\href {\doibase https://doi.org/10.1140/epjc/s10052-019-7206-5} {\bibfield  {journal} {\bibinfo  {journal} {The European Physical Journal C}\ }\textbf {\bibinfo {volume} {79}},\ \bibinfo {pages} {1} (\bibinfo {year} {2019})}\BibitemShut {NoStop}%
\bibitem [{\citenamefont {Moraes}\ \emph {et~al.}(2019)\citenamefont {Moraes}, \citenamefont {De~Paula},\ and\ \citenamefont {Correa}}]{moraes2019charged}%
  \BibitemOpen
  \bibfield  {author} {\bibinfo {author} {\bibfnamefont {P.}~\bibnamefont {Moraes}}, \bibinfo {author} {\bibfnamefont {W.}~\bibnamefont {De~Paula}}, \ and\ \bibinfo {author} {\bibfnamefont {R.}~\bibnamefont {Correa}},\ }\href {\doibase https://doi.org/10.1142/S0218271819500986} {\bibfield  {journal} {\bibinfo  {journal} {International Journal of Modern Physics D}\ }\textbf {\bibinfo {volume} {28}},\ \bibinfo {pages} {1950098} (\bibinfo {year} {2019})}\BibitemShut {NoStop}%
\bibitem [{\citenamefont {Moraes}\ and\ \citenamefont {Sahoo}(2018)}]{moraes2018nonexotic}%
  \BibitemOpen
  \bibfield  {author} {\bibinfo {author} {\bibfnamefont {P.}~\bibnamefont {Moraes}}\ and\ \bibinfo {author} {\bibfnamefont {P.}~\bibnamefont {Sahoo}},\ }\href {\doibase https://doi.org/10.1103/PhysRevD.97.024007} {\bibfield  {journal} {\bibinfo  {journal} {Physical Review D}\ }\textbf {\bibinfo {volume} {97}},\ \bibinfo {pages} {024007} (\bibinfo {year} {2018})}\BibitemShut {NoStop}%
\bibitem [{\citenamefont {Elizalde}\ and\ \citenamefont {Khurshudyan}(2018)}]{elizalde2018wormhole}%
  \BibitemOpen
  \bibfield  {author} {\bibinfo {author} {\bibfnamefont {E.}~\bibnamefont {Elizalde}}\ and\ \bibinfo {author} {\bibfnamefont {M.}~\bibnamefont {Khurshudyan}},\ }\href {\doibase https://doi.org/10.1103/PhysRevD.98.123525} {\bibfield  {journal} {\bibinfo  {journal} {Physical Review D}\ }\textbf {\bibinfo {volume} {98}},\ \bibinfo {pages} {123525} (\bibinfo {year} {2018})}\BibitemShut {NoStop}%
\bibitem [{\citenamefont {Sahoo}\ \emph {et~al.}(2020{\natexlab{a}})\citenamefont {Sahoo}, \citenamefont {Mandal},\ and\ \citenamefont {Sahoo}}]{sahoo2020wormhole}%
  \BibitemOpen
  \bibfield  {author} {\bibinfo {author} {\bibfnamefont {P.}~\bibnamefont {Sahoo}}, \bibinfo {author} {\bibfnamefont {S.}~\bibnamefont {Mandal}}, \ and\ \bibinfo {author} {\bibfnamefont {P.}~\bibnamefont {Sahoo}},\ }\href {\doibase https://doi.org/10.1016/j.newast.2020.101421} {\bibfield  {journal} {\bibinfo  {journal} {New Astronomy}\ }\textbf {\bibinfo {volume} {80}},\ \bibinfo {pages} {101421} (\bibinfo {year} {2020}{\natexlab{a}})}\BibitemShut {NoStop}%
\bibitem [{\citenamefont {Sharif}\ and\ \citenamefont {Siddiqa}(2019)}]{sharif2019propagation}%
  \BibitemOpen
  \bibfield  {author} {\bibinfo {author} {\bibfnamefont {M.}~\bibnamefont {Sharif}}\ and\ \bibinfo {author} {\bibfnamefont {A.}~\bibnamefont {Siddiqa}},\ }\href {\doibase https://doi.org/10.1007/s10714-019-2558-6} {\bibfield  {journal} {\bibinfo  {journal} {General Relativity and Gravitation}\ }\textbf {\bibinfo {volume} {51}},\ \bibinfo {pages} {74} (\bibinfo {year} {2019})}\BibitemShut {NoStop}%
\bibitem [{\citenamefont {Alves}\ \emph {et~al.}(2016)\citenamefont {Alves}, \citenamefont {Moraes}, \citenamefont {De~Araujo},\ and\ \citenamefont {Malheiro}}]{alves2016gravitational}%
  \BibitemOpen
  \bibfield  {author} {\bibinfo {author} {\bibfnamefont {M.}~\bibnamefont {Alves}}, \bibinfo {author} {\bibfnamefont {P.}~\bibnamefont {Moraes}}, \bibinfo {author} {\bibfnamefont {J.}~\bibnamefont {De~Araujo}}, \ and\ \bibinfo {author} {\bibfnamefont {M.}~\bibnamefont {Malheiro}},\ }\href {\doibase https://doi.org/10.1103/PhysRevD.94.024032} {\bibfield  {journal} {\bibinfo  {journal} {Physical Review D}\ }\textbf {\bibinfo {volume} {94}},\ \bibinfo {pages} {024032} (\bibinfo {year} {2016})}\BibitemShut {NoStop}%
\bibitem [{\citenamefont {Sahoo}\ \emph {et~al.}(2020{\natexlab{b}})\citenamefont {Sahoo}, \citenamefont {Bhattacharjee}, \citenamefont {Tripathy},\ and\ \citenamefont {Sahoo}}]{sahoo2020bouncing}%
  \BibitemOpen
  \bibfield  {author} {\bibinfo {author} {\bibfnamefont {P.}~\bibnamefont {Sahoo}}, \bibinfo {author} {\bibfnamefont {S.}~\bibnamefont {Bhattacharjee}}, \bibinfo {author} {\bibfnamefont {S.}~\bibnamefont {Tripathy}}, \ and\ \bibinfo {author} {\bibfnamefont {P.}~\bibnamefont {Sahoo}},\ }\href {\doibase https://doi.org/10.1142/S0217732320500959} {\bibfield  {journal} {\bibinfo  {journal} {Modern Physics Letters A}\ }\textbf {\bibinfo {volume} {35}},\ \bibinfo {pages} {2050095} (\bibinfo {year} {2020}{\natexlab{b}})}\BibitemShut {NoStop}%
\bibitem [{\citenamefont {Bhattacharjee}\ and\ \citenamefont {Sahoo}(2020)}]{bhattacharjee2020redshift}%
  \BibitemOpen
  \bibfield  {author} {\bibinfo {author} {\bibfnamefont {S.}~\bibnamefont {Bhattacharjee}}\ and\ \bibinfo {author} {\bibfnamefont {P.}~\bibnamefont {Sahoo}},\ }\href {\doibase https://doi.org/10.1016/j.newast.2020.101425} {\bibfield  {journal} {\bibinfo  {journal} {New Astronomy}\ }\textbf {\bibinfo {volume} {81}},\ \bibinfo {pages} {101425} (\bibinfo {year} {2020})}\BibitemShut {NoStop}%
\bibitem [{\citenamefont {Sharif}\ and\ \citenamefont {Yousaf}(2014)}]{sharif2014dynamical}%
  \BibitemOpen
  \bibfield  {author} {\bibinfo {author} {\bibfnamefont {M.}~\bibnamefont {Sharif}}\ and\ \bibinfo {author} {\bibfnamefont {Z.}~\bibnamefont {Yousaf}},\ }\href {\doibase https://doi.org/10.1007/s10509-014-2113-6} {\bibfield  {journal} {\bibinfo  {journal} {Astrophysics and space science}\ }\textbf {\bibinfo {volume} {354}},\ \bibinfo {pages} {471} (\bibinfo {year} {2014})}\BibitemShut {NoStop}%
\bibitem [{\citenamefont {Noureen}\ and\ \citenamefont {Zubair}(2015{\natexlab{a}})}]{noureen2015dynamical}%
  \BibitemOpen
  \bibfield  {author} {\bibinfo {author} {\bibfnamefont {I.}~\bibnamefont {Noureen}}\ and\ \bibinfo {author} {\bibfnamefont {M.}~\bibnamefont {Zubair}},\ }\href {\doibase https://doi.org/10.1007/s10509-014-2202-6} {\bibfield  {journal} {\bibinfo  {journal} {Astrophysics and Space Science}\ }\textbf {\bibinfo {volume} {356}},\ \bibinfo {pages} {103} (\bibinfo {year} {2015}{\natexlab{a}})}\BibitemShut {NoStop}%
\bibitem [{\citenamefont {Noureen}\ and\ \citenamefont {Zubair}(2015{\natexlab{b}})}]{noureen2015dynamical1}%
  \BibitemOpen
  \bibfield  {author} {\bibinfo {author} {\bibfnamefont {I.}~\bibnamefont {Noureen}}\ and\ \bibinfo {author} {\bibfnamefont {M.}~\bibnamefont {Zubair}},\ }\href {\doibase https://doi.org/10.1140/epjc/s10052-015-3289-9} {\bibfield  {journal} {\bibinfo  {journal} {The European Physical Journal C}\ }\textbf {\bibinfo {volume} {75}},\ \bibinfo {pages} {62} (\bibinfo {year} {2015}{\natexlab{b}})}\BibitemShut {NoStop}%
\bibitem [{\citenamefont {Zubair}\ \emph {et~al.}(2016)\citenamefont {Zubair}, \citenamefont {Abbas},\ and\ \citenamefont {Noureen}}]{zubair2016possible}%
  \BibitemOpen
  \bibfield  {author} {\bibinfo {author} {\bibfnamefont {M.}~\bibnamefont {Zubair}}, \bibinfo {author} {\bibfnamefont {G.}~\bibnamefont {Abbas}}, \ and\ \bibinfo {author} {\bibfnamefont {I.}~\bibnamefont {Noureen}},\ }\href {\doibase https://doi.org/10.1007/s10509-015-2596-9} {\bibfield  {journal} {\bibinfo  {journal} {Astrophysics and Space Science}\ }\textbf {\bibinfo {volume} {361}},\ \bibinfo {pages} {8} (\bibinfo {year} {2016})}\BibitemShut {NoStop}%
\bibitem [{\citenamefont {Alhamzawi}\ and\ \citenamefont {Alhamzawi}(2016)}]{alhamzawi2016gravitational}%
  \BibitemOpen
  \bibfield  {author} {\bibinfo {author} {\bibfnamefont {A.}~\bibnamefont {Alhamzawi}}\ and\ \bibinfo {author} {\bibfnamefont {R.}~\bibnamefont {Alhamzawi}},\ }\href {\doibase https://doi.org/10.1142/S0218271816500206} {\bibfield  {journal} {\bibinfo  {journal} {International Journal of Modern Physics D}\ }\textbf {\bibinfo {volume} {25}},\ \bibinfo {pages} {1650020} (\bibinfo {year} {2016})}\BibitemShut {NoStop}%
\bibitem [{\citenamefont {Heintzmann}(1969)}]{heintz}%
  \BibitemOpen
  \bibfield  {author} {\bibinfo {author} {\bibfnamefont {H.}~\bibnamefont {Heintzmann}},\ }\href {\doibase 10.1007/BF01558346} {\bibfield  {journal} {\bibinfo  {journal} {Z. Phys., 228: 489-93(1969).}\ }\textbf {\bibinfo {volume} {228}},\ \bibinfo {pages} {489} (\bibinfo {year} {1969})}\BibitemShut {NoStop}%
\bibitem [{\citenamefont {Delgaty}\ and\ \citenamefont {Lake}(1998)}]{Delgaty:1998uy}%
  \BibitemOpen
  \bibfield  {author} {\bibinfo {author} {\bibfnamefont {M.~S.~R.}\ \bibnamefont {Delgaty}}\ and\ \bibinfo {author} {\bibfnamefont {K.}~\bibnamefont {Lake}},\ }\href {\doibase 10.1016/S0010-4655(98)00130-1} {\bibfield  {journal} {\bibinfo  {journal} {Comput. Phys. Commun.}\ }\textbf {\bibinfo {volume} {115}},\ \bibinfo {pages} {395} (\bibinfo {year} {1998})},\ \Eprint {http://arxiv.org/abs/gr-qc/9809013} {arXiv:gr-qc/9809013} \BibitemShut {NoStop}%
\bibitem [{\citenamefont {Andrade}(2022{\natexlab{a}})}]{andrade2022anisotropic}%
  \BibitemOpen
  \bibfield  {author} {\bibinfo {author} {\bibfnamefont {J.}~\bibnamefont {Andrade}},\ }\href {\doibase https://doi.org/10.1140/epjc/s10052-022-10585-6} {\bibfield  {journal} {\bibinfo  {journal} {The European Physical Journal C}\ }\textbf {\bibinfo {volume} {82}},\ \bibinfo {pages} {1} (\bibinfo {year} {2022}{\natexlab{a}})}\BibitemShut {NoStop}%
\bibitem [{\citenamefont {Pradhan}\ and\ \citenamefont {Pant}(2015)}]{pradhan2015anisotropic}%
  \BibitemOpen
  \bibfield  {author} {\bibinfo {author} {\bibfnamefont {N.}~\bibnamefont {Pradhan}}\ and\ \bibinfo {author} {\bibfnamefont {N.}~\bibnamefont {Pant}},\ }\href {\doibase https://doi.org/10.1007/s10509-014-2198-y} {\bibfield  {journal} {\bibinfo  {journal} {Astrophysics and Space Science}\ }\textbf {\bibinfo {volume} {356}},\ \bibinfo {pages} {67} (\bibinfo {year} {2015})}\BibitemShut {NoStop}%
\bibitem [{\citenamefont {Pant}\ \emph {et~al.}(2010)\citenamefont {Pant}, \citenamefont {Mehta},\ and\ \citenamefont {Pant}}]{Pant:2010byj}%
  \BibitemOpen
  \bibfield  {author} {\bibinfo {author} {\bibfnamefont {N.}~\bibnamefont {Pant}}, \bibinfo {author} {\bibfnamefont {R.~N.}\ \bibnamefont {Mehta}}, \ and\ \bibinfo {author} {\bibfnamefont {M.~J.}\ \bibnamefont {Pant}},\ }\href {\doibase 10.1007/s10509-010-0509-5} {\bibfield  {journal} {\bibinfo  {journal} {Astrophys. Space Sci.}\ }\textbf {\bibinfo {volume} {332}},\ \bibinfo {pages} {473} (\bibinfo {year} {2010})}\BibitemShut {NoStop}%
\bibitem [{\citenamefont {Thirukkanesh}\ \emph {et~al.}(2018)\citenamefont {Thirukkanesh}, \citenamefont {Ragel}, \citenamefont {Sharma},\ and\ \citenamefont {Das}}]{Thirukkanesh:2018hfy}%
  \BibitemOpen
  \bibfield  {author} {\bibinfo {author} {\bibfnamefont {S.}~\bibnamefont {Thirukkanesh}}, \bibinfo {author} {\bibfnamefont {F.~C.}\ \bibnamefont {Ragel}}, \bibinfo {author} {\bibfnamefont {R.}~\bibnamefont {Sharma}}, \ and\ \bibinfo {author} {\bibfnamefont {S.}~\bibnamefont {Das}},\ }\href {\doibase 10.1140/epjc/s10052-018-5526-5} {\bibfield  {journal} {\bibinfo  {journal} {Eur. Phys. J. C}\ }\textbf {\bibinfo {volume} {78}},\ \bibinfo {pages} {31} (\bibinfo {year} {2018})},\ \Eprint {http://arxiv.org/abs/1801.02956} {arXiv:1801.02956 [physics.gen-ph]} \BibitemShut {NoStop}%
\bibitem [{\citenamefont {Singh}\ and\ \citenamefont {Pant}(2016)}]{Singh:2015kyr}%
  \BibitemOpen
  \bibfield  {author} {\bibinfo {author} {\bibfnamefont {K.~N.}\ \bibnamefont {Singh}}\ and\ \bibinfo {author} {\bibfnamefont {N.}~\bibnamefont {Pant}},\ }\href {\doibase 10.1007/s12648-015-0815-4} {\bibfield  {journal} {\bibinfo  {journal} {Indian J. Phys.}\ }\textbf {\bibinfo {volume} {90}},\ \bibinfo {pages} {843} (\bibinfo {year} {2016})}\BibitemShut {NoStop}%
\bibitem [{\citenamefont {Estrada}\ and\ \citenamefont {Tello-Ortiz}(2018)}]{Estrada:2018zbh}%
  \BibitemOpen
  \bibfield  {author} {\bibinfo {author} {\bibfnamefont {M.}~\bibnamefont {Estrada}}\ and\ \bibinfo {author} {\bibfnamefont {F.}~\bibnamefont {Tello-Ortiz}},\ }\href {\doibase 10.1140/epjp/i2018-12249-9} {\bibfield  {journal} {\bibinfo  {journal} {Eur. Phys. J. Plus}\ }\textbf {\bibinfo {volume} {133}},\ \bibinfo {pages} {453} (\bibinfo {year} {2018})},\ \Eprint {http://arxiv.org/abs/1803.02344} {arXiv:1803.02344 [gr-qc]} \BibitemShut {NoStop}%
\bibitem [{\citenamefont {Morales}\ and\ \citenamefont {Tello-Ortiz}(2018)}]{Morales:2018nmq}%
  \BibitemOpen
  \bibfield  {author} {\bibinfo {author} {\bibfnamefont {E.}~\bibnamefont {Morales}}\ and\ \bibinfo {author} {\bibfnamefont {F.}~\bibnamefont {Tello-Ortiz}},\ }\href {\doibase 10.1140/epjc/s10052-018-6102-8} {\bibfield  {journal} {\bibinfo  {journal} {Eur. Phys. J. C}\ }\textbf {\bibinfo {volume} {78}},\ \bibinfo {pages} {618} (\bibinfo {year} {2018})},\ \Eprint {http://arxiv.org/abs/1805.00592} {arXiv:1805.00592 [gr-qc]} \BibitemShut {NoStop}%
\bibitem [{\citenamefont {Zubair}\ \emph {et~al.}(2021)\citenamefont {Zubair}, \citenamefont {Amin},\ and\ \citenamefont {Azmat}}]{Zubair:2021zqs}%
  \BibitemOpen
  \bibfield  {author} {\bibinfo {author} {\bibfnamefont {M.}~\bibnamefont {Zubair}}, \bibinfo {author} {\bibfnamefont {M.}~\bibnamefont {Amin}}, \ and\ \bibinfo {author} {\bibfnamefont {H.}~\bibnamefont {Azmat}},\ }\href {\doibase 10.1088/1402-4896/ac237d} {\bibfield  {journal} {\bibinfo  {journal} {Phys. Scripta}\ }\textbf {\bibinfo {volume} {96}} (\bibinfo {year} {2021}),\ 10.1088/1402-4896/ac237d}\BibitemShut {NoStop}%
\bibitem [{\citenamefont {Sharif}\ and\ \citenamefont {Majid}(2020)}]{Sharif:2020lbt}%
  \BibitemOpen
  \bibfield  {author} {\bibinfo {author} {\bibfnamefont {M.}~\bibnamefont {Sharif}}\ and\ \bibinfo {author} {\bibfnamefont {A.}~\bibnamefont {Majid}},\ }\href {\doibase 10.1016/j.cjph.2020.09.015} {\bibfield  {journal} {\bibinfo  {journal} {Chin. J. Phys.}\ }\textbf {\bibinfo {volume} {68}},\ \bibinfo {pages} {406} (\bibinfo {year} {2020})}\BibitemShut {NoStop}%
\bibitem [{\citenamefont {Sharif}\ and\ \citenamefont {Waseem}(2020)}]{Sharif:2019zyh}%
  \BibitemOpen
  \bibfield  {author} {\bibinfo {author} {\bibfnamefont {M.}~\bibnamefont {Sharif}}\ and\ \bibinfo {author} {\bibfnamefont {A.}~\bibnamefont {Waseem}},\ }\href {\doibase 10.1016/j.cjph.2019.11.006} {\bibfield  {journal} {\bibinfo  {journal} {Chin. J. Phys.}\ }\textbf {\bibinfo {volume} {63}},\ \bibinfo {pages} {92} (\bibinfo {year} {2020})},\ \Eprint {http://arxiv.org/abs/1912.06480} {arXiv:1912.06480 [gr-qc]} \BibitemShut {NoStop}%
\bibitem [{\citenamefont {Andrade}(2022{\natexlab{b}})}]{Andrade:2022idg}%
  \BibitemOpen
  \bibfield  {author} {\bibinfo {author} {\bibfnamefont {J.}~\bibnamefont {Andrade}},\ }\href {\doibase 10.1140/epjc/s10052-022-10585-6} {\bibfield  {journal} {\bibinfo  {journal} {Eur. Phys. J. C}\ }\textbf {\bibinfo {volume} {82}},\ \bibinfo {pages} {617} (\bibinfo {year} {2022}{\natexlab{b}})}\BibitemShut {NoStop}%
\bibitem [{\citenamefont {Rawls}\ \emph {et~al.}(2011)\citenamefont {Rawls}, \citenamefont {Orosz}, \citenamefont {McClintock}, \citenamefont {Torres}, \citenamefont {Bailyn},\ and\ \citenamefont {Buxton}}]{Rawls:2011jw}%
  \BibitemOpen
  \bibfield  {author} {\bibinfo {author} {\bibfnamefont {M.~L.}\ \bibnamefont {Rawls}}, \bibinfo {author} {\bibfnamefont {J.~A.}\ \bibnamefont {Orosz}}, \bibinfo {author} {\bibfnamefont {J.~E.}\ \bibnamefont {McClintock}}, \bibinfo {author} {\bibfnamefont {M.~A.~P.}\ \bibnamefont {Torres}}, \bibinfo {author} {\bibfnamefont {C.~D.}\ \bibnamefont {Bailyn}}, \ and\ \bibinfo {author} {\bibfnamefont {M.~M.}\ \bibnamefont {Buxton}},\ }\href {\doibase 10.1088/0004-637X/730/1/25} {\bibfield  {journal} {\bibinfo  {journal} {Astrophys. J.}\ }\textbf {\bibinfo {volume} {730}},\ \bibinfo {pages} {25} (\bibinfo {year} {2011})},\ \Eprint {http://arxiv.org/abs/1101.2465} {arXiv:1101.2465 [astro-ph.SR]} \BibitemShut {NoStop}%
\bibitem [{\citenamefont {Sako}\ \emph {et~al.}(2000)\citenamefont {Sako}, \citenamefont {Kahn}, \citenamefont {Paerels},\ and\ \citenamefont {Liedahl}}]{Sako:2000ve}%
  \BibitemOpen
  \bibfield  {author} {\bibinfo {author} {\bibfnamefont {M.}~\bibnamefont {Sako}}, \bibinfo {author} {\bibfnamefont {S.~M.}\ \bibnamefont {Kahn}}, \bibinfo {author} {\bibfnamefont {F.}~\bibnamefont {Paerels}}, \ and\ \bibinfo {author} {\bibfnamefont {D.~A.}\ \bibnamefont {Liedahl}},\ }\href {\doibase 10.1086/317053} {\bibfield  {journal} {\bibinfo  {journal} {Astrophys. J.}\ }\textbf {\bibinfo {volume} {542}},\ \bibinfo {pages} {684} (\bibinfo {year} {2000})},\ \Eprint {http://arxiv.org/abs/astro-ph/0006146} {arXiv:astro-ph/0006146} \BibitemShut {NoStop}%
\bibitem [{\citenamefont {Landau}(2013)}]{landau2013classical}%
  \BibitemOpen
  \bibfield  {author} {\bibinfo {author} {\bibfnamefont {L.~D.}\ \bibnamefont {Landau}},\ }\href {https://books.google.co.in/books?id=HudbAwAAQBAJ&dq=The+classical+theory+of+fields%7D,+++author%3D%7BLandau,+Lev+Davidovich%7D&lr=&source=gbs_navlinks_s} {\emph {\bibinfo {title} {The classical theory of fields}}},\ Vol.~\bibinfo {volume} {2}\ (\bibinfo  {publisher} {Elsevier},\ \bibinfo {year} {2013})\BibitemShut {NoStop}%
\bibitem [{\citenamefont {Koivisto}(2006)}]{koivisto2006note}%
  \BibitemOpen
  \bibfield  {author} {\bibinfo {author} {\bibfnamefont {T.}~\bibnamefont {Koivisto}},\ }\href {\doibase 10.1088/0264-9381/23/12/N01} {\bibfield  {journal} {\bibinfo  {journal} {Classical and Quantum Gravity}\ }\textbf {\bibinfo {volume} {23}},\ \bibinfo {pages} {4289} (\bibinfo {year} {2006})}\BibitemShut {NoStop}%
\bibitem [{\citenamefont {Lobato}\ \emph {et~al.}(2019)\citenamefont {Lobato}, \citenamefont {Carvalho}, \citenamefont {Martins},\ and\ \citenamefont {Moraes}}]{Lobato:2018vpq}%
  \BibitemOpen
  \bibfield  {author} {\bibinfo {author} {\bibfnamefont {R.~V.}\ \bibnamefont {Lobato}}, \bibinfo {author} {\bibfnamefont {G.~A.}\ \bibnamefont {Carvalho}}, \bibinfo {author} {\bibfnamefont {A.~G.}\ \bibnamefont {Martins}}, \ and\ \bibinfo {author} {\bibfnamefont {P.~H. R.~S.}\ \bibnamefont {Moraes}},\ }\href {\doibase 10.1140/epjp/i2019-12638-6} {\bibfield  {journal} {\bibinfo  {journal} {Eur. Phys. J. Plus}\ }\textbf {\bibinfo {volume} {134}},\ \bibinfo {pages} {132} (\bibinfo {year} {2019})},\ \Eprint {http://arxiv.org/abs/1803.08630} {arXiv:1803.08630 [gr-qc]} \BibitemShut {NoStop}%
\bibitem [{\citenamefont {Pretel}\ \emph {et~al.}(2021)\citenamefont {Pretel}, \citenamefont {Jor\'as}, \citenamefont {Reis},\ and\ \citenamefont {Arba\~nil}}]{Pretel:2021kgl}%
  \BibitemOpen
  \bibfield  {author} {\bibinfo {author} {\bibfnamefont {J.~M.~Z.}\ \bibnamefont {Pretel}}, \bibinfo {author} {\bibfnamefont {S.~E.}\ \bibnamefont {Jor\'as}}, \bibinfo {author} {\bibfnamefont {R.~R.~R.}\ \bibnamefont {Reis}}, \ and\ \bibinfo {author} {\bibfnamefont {J.~D.~V.}\ \bibnamefont {Arba\~nil}},\ }\href {\doibase 10.1088/1475-7516/2021/08/055} {\bibfield  {journal} {\bibinfo  {journal} {JCAP}\ }\textbf {\bibinfo {volume} {08}},\ \bibinfo {pages} {055} (\bibinfo {year} {2021})},\ \Eprint {http://arxiv.org/abs/2105.07573} {arXiv:2105.07573 [gr-qc]} \BibitemShut {NoStop}%
\bibitem [{\citenamefont {Carvalho}\ \emph {et~al.}(2020)\citenamefont {Carvalho}, \citenamefont {Dos~Santos}, \citenamefont {Moraes},\ and\ \citenamefont {Malheiro}}]{Carvalho:2019gzs}%
  \BibitemOpen
  \bibfield  {author} {\bibinfo {author} {\bibfnamefont {G.~A.}\ \bibnamefont {Carvalho}}, \bibinfo {author} {\bibfnamefont {S.~I.}\ \bibnamefont {Dos~Santos}}, \bibinfo {author} {\bibfnamefont {P.~H. R.~S.}\ \bibnamefont {Moraes}}, \ and\ \bibinfo {author} {\bibfnamefont {M.}~\bibnamefont {Malheiro}},\ }\href {\doibase 10.1142/S0218271820500753} {\bibfield  {journal} {\bibinfo  {journal} {Int. J. Mod. Phys. D}\ }\textbf {\bibinfo {volume} {29}},\ \bibinfo {pages} {2050075} (\bibinfo {year} {2020})},\ \Eprint {http://arxiv.org/abs/1911.02484} {arXiv:1911.02484 [gr-qc]} \BibitemShut {NoStop}%
\bibitem [{\citenamefont {Sharifa}\ and\ \citenamefont {Waseemb}(2018)}]{sharifa2018anisotropic}%
  \BibitemOpen
  \bibfield  {author} {\bibinfo {author} {\bibfnamefont {M.}~\bibnamefont {Sharifa}}\ and\ \bibinfo {author} {\bibfnamefont {A.}~\bibnamefont {Waseemb}},\ }\href {\doibase https://doi.org/10.1140/epjc/s10052-018-6363-2} {\bibfield  {journal} {\bibinfo  {journal} {Eur. Phys. J. C}\ }\textbf {\bibinfo {volume} {78}},\ \bibinfo {pages} {868} (\bibinfo {year} {2018})}\BibitemShut {NoStop}%
\bibitem [{\citenamefont {Kokkotas}\ and\ \citenamefont {Ruoff}(2001)}]{kokkotas2001radial}%
  \BibitemOpen
  \bibfield  {author} {\bibinfo {author} {\bibfnamefont {K.}~\bibnamefont {Kokkotas}}\ and\ \bibinfo {author} {\bibfnamefont {J.}~\bibnamefont {Ruoff}},\ }\href {\doibase 10.1051/0004-6361:20000216} {\bibfield  {journal} {\bibinfo  {journal} {Astronomy \& Astrophysics}\ }\textbf {\bibinfo {volume} {366}},\ \bibinfo {pages} {565} (\bibinfo {year} {2001})}\BibitemShut {NoStop}%
\bibitem [{\citenamefont {Chandrasekhar}(1964)}]{Chandrasekhar:1964zz}%
  \BibitemOpen
  \bibfield  {author} {\bibinfo {author} {\bibfnamefont {S.}~\bibnamefont {Chandrasekhar}},\ }\href {\doibase 10.1086/147938} {\bibfield  {journal} {\bibinfo  {journal} {Astrophys. J.}\ }\textbf {\bibinfo {volume} {140}},\ \bibinfo {pages} {417} (\bibinfo {year} {1964})},\ \bibinfo {note} {[Erratum: Astrophys.J. 140, 1342 (1964)]}\BibitemShut {NoStop}%
\bibitem [{\citenamefont {Heintzmann}\ and\ \citenamefont {Hillebrandt}(1975)}]{heintzmann1975neutron}%
  \BibitemOpen
  \bibfield  {author} {\bibinfo {author} {\bibfnamefont {H.}~\bibnamefont {Heintzmann}}\ and\ \bibinfo {author} {\bibfnamefont {W.}~\bibnamefont {Hillebrandt}},\ }\href {https://ui.adsabs.harvard.edu/abs/1975A&A....38...51H} {\bibfield  {journal} {\bibinfo  {journal} {Astronomy and Astrophysics}\ }\textbf {\bibinfo {volume} {38}},\ \bibinfo {pages} {51} (\bibinfo {year} {1975})}\BibitemShut {NoStop}%
\bibitem [{\citenamefont {Hillebrandt}\ and\ \citenamefont {Steinmetz}(1976)}]{hillebrandt1976anisotropic}%
  \BibitemOpen
  \bibfield  {author} {\bibinfo {author} {\bibfnamefont {W.}~\bibnamefont {Hillebrandt}}\ and\ \bibinfo {author} {\bibfnamefont {K.}~\bibnamefont {Steinmetz}},\ }\href {https://adsabs.harvard.edu/full/1976A%26A....53..283H} {\bibfield  {journal} {\bibinfo  {journal} {Astronomy and Astrophysics}\ }\textbf {\bibinfo {volume} {53}},\ \bibinfo {pages} {283} (\bibinfo {year} {1976})}\BibitemShut {NoStop}%
\bibitem [{\citenamefont {Bhar}\ \emph {et~al.}(2019)\citenamefont {Bhar}, \citenamefont {Singh},\ and\ \citenamefont {Tello-Ortiz}}]{bhar2019compact}%
  \BibitemOpen
  \bibfield  {author} {\bibinfo {author} {\bibfnamefont {P.}~\bibnamefont {Bhar}}, \bibinfo {author} {\bibfnamefont {K.~N.}\ \bibnamefont {Singh}}, \ and\ \bibinfo {author} {\bibfnamefont {F.}~\bibnamefont {Tello-Ortiz}},\ }\href {\doibase https://doi.org/10.1140/epjc/s10052-019-7438-4} {\bibfield  {journal} {\bibinfo  {journal} {The European Physical Journal C}\ }\textbf {\bibinfo {volume} {79}},\ \bibinfo {pages} {922} (\bibinfo {year} {2019})}\BibitemShut {NoStop}%
\bibitem [{\citenamefont {Jasim}\ \emph {et~al.}(2021)\citenamefont {Jasim}, \citenamefont {Maurya}, \citenamefont {Singh},\ and\ \citenamefont {Nag}}]{jasim2021anisotropic}%
  \BibitemOpen
  \bibfield  {author} {\bibinfo {author} {\bibfnamefont {M.~K.}\ \bibnamefont {Jasim}}, \bibinfo {author} {\bibfnamefont {S.~K.}\ \bibnamefont {Maurya}}, \bibinfo {author} {\bibfnamefont {K.~N.}\ \bibnamefont {Singh}}, \ and\ \bibinfo {author} {\bibfnamefont {R.}~\bibnamefont {Nag}},\ }\href {\doibase 1015; https://doi.org/10.3390/e23081015} {\bibfield  {journal} {\bibinfo  {journal} {Entropy}\ }\textbf {\bibinfo {volume} {23}},\ \bibinfo {pages} {1015} (\bibinfo {year} {2021})}\BibitemShut {NoStop}%
\bibitem [{\citenamefont {Zel'dovich}(1962)}]{l1962equation}%
  \BibitemOpen
  \bibfield  {author} {\bibinfo {author} {\bibfnamefont {Y.~B.}\ \bibnamefont {Zel'dovich}},\ }\href {http://www.jetp.ras.ru/cgi-bin/e/index/e/14/5/p1143?a=list} {\bibfield  {journal} {\bibinfo  {journal} {Soviet physics JETP}\ }\textbf {\bibinfo {volume} {14}} (\bibinfo {year} {1962})}\BibitemShut {NoStop}%
\bibitem [{\citenamefont {Zeldovich}\ and\ \citenamefont {Novikov}(1971)}]{zeldovich1971relativistic}%
  \BibitemOpen
  \bibfield  {author} {\bibinfo {author} {\bibfnamefont {Y.~B.}\ \bibnamefont {Zeldovich}}\ and\ \bibinfo {author} {\bibfnamefont {I.~D.}\ \bibnamefont {Novikov}},\ }\href {https://doi.org/10.1063/1.3070777} {\bibfield  {journal} {\bibinfo  {journal} {Chicago: University of Chicago Press}\ } (\bibinfo {year} {1971})}\BibitemShut {NoStop}%
\bibitem [{\citenamefont {Das}\ \emph {et~al.}(2016)\citenamefont {Das}, \citenamefont {Rahaman}, \citenamefont {Guha},\ and\ \citenamefont {Ray}}]{das2016compact}%
  \BibitemOpen
  \bibfield  {author} {\bibinfo {author} {\bibfnamefont {A.}~\bibnamefont {Das}}, \bibinfo {author} {\bibfnamefont {F.}~\bibnamefont {Rahaman}}, \bibinfo {author} {\bibfnamefont {B.}~\bibnamefont {Guha}}, \ and\ \bibinfo {author} {\bibfnamefont {S.}~\bibnamefont {Ray}},\ }\href {\doibase DOI 10.1140/epjc/s10052-016-4503-0} {\bibfield  {journal} {\bibinfo  {journal} {The European Physical Journal C}\ }\textbf {\bibinfo {volume} {76}},\ \bibinfo {pages} {1} (\bibinfo {year} {2016})}\BibitemShut {NoStop}%
\bibitem [{\citenamefont {Harrison}\ \emph {et~al.}(1965)\citenamefont {Harrison}, \citenamefont {Thorne}, \citenamefont {Wakano},\ and\ \citenamefont {Wheeler}}]{harrison1965gravitation}%
  \BibitemOpen
  \bibfield  {author} {\bibinfo {author} {\bibfnamefont {B.~K.}\ \bibnamefont {Harrison}}, \bibinfo {author} {\bibfnamefont {K.~S.}\ \bibnamefont {Thorne}}, \bibinfo {author} {\bibfnamefont {M.}~\bibnamefont {Wakano}}, \ and\ \bibinfo {author} {\bibfnamefont {J.~A.}\ \bibnamefont {Wheeler}},\ }\href@noop {} {\bibfield  {journal} {\bibinfo  {journal} {Gravitation Theory and Gravitational Collapse}\ } (\bibinfo {year} {1965})}\BibitemShut {NoStop}%
\bibitem [{\citenamefont {Glendenning}\ and\ \citenamefont {Kettner}(2000)}]{Glendenning:1998ag}%
  \BibitemOpen
  \bibfield  {author} {\bibinfo {author} {\bibfnamefont {N.~K.}\ \bibnamefont {Glendenning}}\ and\ \bibinfo {author} {\bibfnamefont {C.}~\bibnamefont {Kettner}},\ }\href@noop {} {\bibfield  {journal} {\bibinfo  {journal} {Astron. Astrophys.}\ }\textbf {\bibinfo {volume} {353}},\ \bibinfo {pages} {L9} (\bibinfo {year} {2000})},\ \Eprint {http://arxiv.org/abs/astro-ph/9807155} {arXiv:astro-ph/9807155} \BibitemShut {NoStop}%
\bibitem [{\citenamefont {Arba\~nil}\ and\ \citenamefont {Malheiro}(2016)}]{Arbanil:2016wud}%
  \BibitemOpen
  \bibfield  {author} {\bibinfo {author} {\bibfnamefont {J.~D.~V.}\ \bibnamefont {Arba\~nil}}\ and\ \bibinfo {author} {\bibfnamefont {M.}~\bibnamefont {Malheiro}},\ }\href {\doibase 10.1088/1475-7516/2016/11/012} {\bibfield  {journal} {\bibinfo  {journal} {JCAP}\ }\textbf {\bibinfo {volume} {11}},\ \bibinfo {pages} {012} (\bibinfo {year} {2016})},\ \Eprint {http://arxiv.org/abs/1607.03984} {arXiv:1607.03984 [astro-ph.HE]} \BibitemShut {NoStop}%
\end{thebibliography}%

\end{document}